\documentclass[12pt,letterpaper]{article}
\pdfoutput=1

\usepackage{amsmath,amssymb,calc, amsthm,bbm, epsfig,psfrag, mathtools,comment}

\usepackage{mathrsfs}
\usepackage{physics}
\usepackage{graphicx, enumerate}
\usepackage{float}

\usepackage[numbers,sort&compress]{natbib}

\usepackage[dvipsnames]{xcolor}
\usepackage{tikz}

\usepackage{quiver}
\usepackage{tikz-cd}
\usepackage{enumitem}
\usepackage{tensor}

\makeatletter
\newcommand*\bigcdot{\mathpalette\bigcdot@{.5}}
\newcommand*\bigcdot@[2]{\mathbin{\vcenter{\hbox{\scalebox{#2}{$\m@th#1\bullet$}}}}}
\makeatother

\newcommand{\deq}{\stackrel{\bigcdot}{=}}
\newcommand{\dapprox}{\stackrel{\bigcdot}{\approx}}

\usepackage[utf8]{inputenc} 
\definecolor{azure}{rgb}{0.0, 0.5, 1.0}
\definecolor{darkblue}{rgb}{0.15,0.35,0.7}
\definecolor{reddish}{rgb}{0.65, 0.2, 0.2}
\definecolor{brandeisblue}{rgb}{0.0, 0.44, 1.0}
\definecolor{ceruleanblue}{rgb}{0.16, 0.32, 0.75}
\definecolor{indigo(dye)}{rgb}{0.0, 0.25, 0.42}
\usepackage[linktocpage=true]{hyperref}
\hypersetup{
colorlinks=true,
citecolor=ceruleanblue,
linkcolor=ceruleanblue,
urlcolor=ceruleanblue,
pdfauthor={},
pdftitle={},
pdfsubject={}
}

\newcommand{\overbar}[1]{\mkern 1.5mu\overline{\mkern-1.5mu#1\mkern-1.5mu}\mkern 1.5mu}

\newcommand{\TT}{T\overbar{T}}

\newcommand\Rg{{\mathcal{R}_g}}

\newcommand{\HJ}{\widehat{J}}
\newcommand{\HCCJ}{\widehat{\mathfrak{J}}}

\DeclareFontEncoding{LS1}{}{}
\DeclareFontSubstitution{LS1}{stix}{m}{n}
\DeclareSymbolFont{stixsymbols}{LS1}{stixscr}{m}{n}
\SetSymbolFont{stixsymbols}{bold}{LS1}{stixscr}{b}{n}
\DeclareMathSymbol{\kay}{\mathalpha}{stixsymbols}{"6B}
\DeclareMathSymbol{\hay}{\mathalpha}{stixsymbols}{"68}

\newcommand\CC[1]{\mathfrak{#1}}

\def\bea{\begin{eqnarray}}
\def\eea{\end{eqnarray}}

\makeatletter
\renewcommand\section{\@startsection {section}{1}{\z@}%
                               {-3.5ex \@plus -1ex \@minus -.2ex}
                               {2.3ex \@plus.2ex}%
                               {\normalfont\large\bfseries}}
\renewcommand\subsection{\@startsection{subsection}{2}{\z@}%
                                 {-3.25ex\@plus -1ex \@minus -.2ex}%
                                 {1.5ex \@plus .2ex}%
                                 {\normalfont\bfseries}}
\makeatother

\newfont{\goth}{ygoth.tfm scaled 1200}                   

\numberwithin{equation}{section}

\begin{document}
\begin{titlepage}
\begin{flushright}
\today
\end{flushright}
\vspace{5mm}

\begin{center}
{\Large \bf 
Auxiliary Field Sigma Models \\ and Yang-Baxter Deformations}
\end{center}

\begin{center}

{\bf
Daniele Bielli${}^{a}$,
Christian Ferko${}^{b}$,
Liam Smith${}^{c}$,\\
Gabriele Tartaglino-Mazzucchelli${}^{c}$
} \\
\vspace{5mm}

\footnotesize{
${}^{a}$
{\it 
High Energy Physics Research Unit, Faculty of Science \\ 
Chulalongkorn University, Bangkok 10330, Thailand
}
 \\~\\
${}^{b}$
{\it 
Center for Quantum Mathematics and Physics (QMAP), 
\\ Department of Physics \& Astronomy,  University of California, Davis, CA 95616, USA
}
 \\~\\
${}^{c}$
{\it 
School of Mathematics and Physics, University of Queensland,
\\
 St Lucia, Brisbane, Queensland 4072, Australia}
}
\vspace{2mm}
~\\
\texttt{d.bielli4@gmail.com,
caferko@ucdavis.edu,
liam.smith1@uq.net.au,
g.tartaglino-mazzucchelli@uq.edu.au
}\\
\vspace{2mm}

\end{center}

\begin{abstract}
\baselineskip=14pt

We combine the Yang-Baxter (YB) and bi-Yang-Baxter (bi-YB) deformations with higher-spin auxiliary field deformations 
to construct multi-parameter families of integrable deformations of the principal chiral model on a Lie group $G$ with semi-simple Lie algebra $\mathfrak{g}$. In the YB case, our construction produces one integrable deformation for each pair $(\mathcal{R}, E)$, where $\mathcal{R}$ is an antisymmetric bilinear operator on $\mathfrak{g}$ obeying the modified classical Yang-Baxter equation and $E$ is a function of several variables. In the bi-YB case, the pair becomes a triplet $(\mathcal{R},\tilde{\mathcal{R}}, E)$, where $\tilde{\mathcal{R}}$ is another antisymmetric bilinear operator on $\mathfrak{g}$ and obeys the non-split inhomogeneous modified classical Yang-Baxter equation.  We show that every model in these families is (weakly) classically integrable by exhibiting a Lax representation for their equations of motion.

\end{abstract}
\vspace{5mm}

\vfill
\end{titlepage}

\tableofcontents
\bigskip\hrule

\section{Introduction}

Although integrable quantum field theories (IQFTs) in two spacetime dimensions have been studied since the 1970s, this subject has attracted renewed interest in recent years due to the appearance of integrable structures in string theory and holography. Famously, the Green-Schwarz action which describes superstring propagation on a spacetime with manifest target-space supersymmetry can -- in cases where the target space admits a supercoset structure, which includes many scenarios of interest -- be presented using a Metsaev-Tseytlin construction \cite{Metsaev:1998it,Bena:2003wd}, from which one can show that the worldsheet theory is classically integrable. The integrability of the $2d$ worldsheet theory is a powerful tool which has allowed exact computation of physical quantities, such as the spectrum, for strings at finite tension propagating on curved spacetime manifolds -- see, for instance, the reviews \cite{Beisert:2010jr,Demulder:2023bux}.

The utility of integrable structures in worldsheet string theory has provided additional motivation to search for deformations of $2d$ IQFTs which preserve integrability. In particular, for applications to string theory, it is most interesting to identify integrable deformations of $2d$ sigma models. Perhaps the simplest example of such an integrable sigma model is the principal chiral model (PCM), whose Lagrangian takes the form
\begin{align}\label{pcm_defn}
    \mathcal{L}_{\text{PCM}} = - \frac{1}{2} \tr \left( j_+  j_- \right) \, ,
\end{align}
when written in light-cone coordinates on the $2d$ worldsheet $\Sigma$. The fundamental field of this model is a group-valued field $g : \Sigma \to G$, where $G$ is a Lie group with Lie algebra $\mathfrak{g}$, and the object $j_{\pm}$ appearing in (\ref{pcm_defn}) is the left-invariant Maurer-Cartan form,
\begin{align}
    j_{\pm} = g^{-1} \partial_{\pm} g \, ,
\end{align}
that takes values in $\mathfrak{g}$, which we always assume to be semi-simple.

In this work, we will discuss two particular integrable deformations of the $2d$ principal chiral model, namely Yang-Baxter (YB)  deformations and auxiliary field deformations. Both of these deformations might be described as somewhat ``universal'' in the sense that they can be applied to the PCM associated with any Lie group $G$, and also to more general models, as we will explain shortly. This universality means that both of these types of deformations can be viewed as systematic procedures for generating families of integrable models beginning from an integrable ``seed'' theory; in our case, the role of the seed is played by the PCM for a given choice of Lie group $G$. It is natural to ask how these two deformations interact and whether they can be simultaneously activated.

Let us briefly review each of the two integrable deformations mentioned above. The first of these, the Yang-Baxter deformation, was introduced as an integrable deformation of the PCM by Klim\v{c}\'{i}k \cite{Klimcik:2002zj,Klimcik:2008eq}.\footnote{\label{biyb_foot}The Yang-Baxter deformation was further extended to the so-called bi-Yang-Baxter (bi-YB) deformation in \cite{Klimcik:2014bta}. To keep the presentation simple, we refrain from reviewing the bi-Yang-Baxter case in detail here. However, the reader should keep in mind that as part of the results in our work, we will in fact extend bi-Yang-Baxter to include arbitrary higher-spin auxiliary field deformations.} It is sometimes also called the $\eta$-deformation due to the parameter $\eta$ that appears in the Lagrangian, and where the undeformed PCM is recovered in the limit where $\eta$ is taken to zero.\footnote{For a pedagogical introduction to Yang-Baxter deformations in $2d$ sigma models, see the textbook \cite{yoshida2021yang}.} The data which defines a Yang-Baxter deformation is a linear operator $\mathcal{R} : \mathfrak{g} \to \mathfrak{g}$ acting on the Lie algebra $\mathfrak{g}$ of $G$, which is antisymmetric and solves the modified classical Yang-Baxter equation (mCYBE):
\begin{align}\label{mcybe_intro}
    [ \mathcal{R} X , \mathcal{R} Y ] - \mathcal{R} \left( [ X , \mathcal{R} Y ] \right) - \mathcal{R} \left( [ \mathcal{R} X , Y ] \right) + c^2 [ X, Y ] = 0 \, , \qquad \forall \, X,Y \in \mathfrak{g} \,\, .
\end{align}
The term ``modified'' refers to the case of (\ref{mcybe_intro}) where $c \neq 0$, while when $c = 0$ this equation is referred to as the classical Yang-Baxter equation (CYBE), and the corresponding Yang-Baxter deformation is said to be homogeneous. When $c \neq 0$, one further distinguishes between the ``split inhomogeneous case'' when $c^2 > 0$ and the ``non-split inhomogeneous case'' when $c^2 < 0$. The latter, non-split case is the one which was considered in the original work of Klim\v{c}\'{i}k. In these two distinct inhomogeneous cases, by a rescaling of the linear operator $\mathcal{R}$, one can always choose to take $c^2 = 1$ or $c^2 = -1$, respectively.
    
Inhomogeneous Yang-Baxter deformations of the PCM were subsequently generalized to deformations of symmetric and semi-symmetric space sigma models \cite{Delduc:2013fga,Delduc:2013qra,Delduc:2014kha}. These are precisely the models of interest for studying type IIB string propagation on target spacetimes such as $\mathrm{AdS}_5 \times S^5$, so it is natural to ask whether a Yang-Baxter deformation of the worldsheet theory in this setting is associated with a new solution to the equations of motion for type IIB supergravity \cite{Arutyunov:2015qva}. This line of inquiry spurred the development of ``generalized supergravity'' \cite{Arutyunov:2015mqj,Wulff:2016tju}. In short, the Yang-Baxter deformation corresponds to a background which is a solution of type IIB supergravity if the classical $r$-matrix associated with the deformation satisfies a constraint known as the unimodularity condition \cite{Borsato:2016ose,Hoare:2018ngg}, and otherwise, the background is instead a solution to generalized supergravity equations.

The case $c = 0$, corresponding to homogeneous Yang-Baxter (hYB) deformations, is also of considerable interest following a few of the seminal works \cite{Kawaguchi:2014qwa,Matsumoto:2015jja,vanTongeren:2015soa}.\footnote{The homogeneous Yang-Baxter deformation can be applied to any $2d$ sigma model admitting a Lie group $G$ of isometries \cite{Bakhmatov:2017joy,Bakhmatov:2018apn,Borsato:2018idb}, not just the PCM, which is another sense in which it is ``universal.''} As in the case of inhomogeneous Yang-Baxter deformations, where an investigation of the target-space properties of the deformation led to fruitful results on generalized supergravity, it is natural to wonder about a target-space interpretation in the homogeneous case. It turns out that Abelian homogeneous YB deformations (for which the associated generators are all commuting) are equivalent to sequences of TsT or ``T-duality, shift, T-duality'' transformations in the target spacetime \cite{Matsumoto:2014nra,vanTongeren:2015soa,Osten:2016dvf}. This observation has also been extended via generalizations which relate homogeneous Yang-Baxter deformations to non-Abelian T-duality \cite{Hoare:2016wsk,Borsato:2016pas,Borsato:2017qsx,Borsato:2018idb}. Finally, hYB deformations have also been recast as a presentation of the undeformed theory which is subject to twisted boundary conditions \cite{Borsato:2021fuy}.

Given the wealth of interesting results and connections that have arisen from the study of YB deformations, it is desirable to combine these deformations with other integrable deformations of sigma models to generate larger families of models which may likewise lead to further insights. This has already been accomplished in several cases, such as the combination of YB deformations of the PCM with a Wess-Zumino (WZ) term \cite{Hoare:2020mpv}, YB deformations of WZW models based on Lie supergroups \cite{Eghbali:2022lqn}, a generalization called the bi-Yang-Baxter deformation \cite{Klimcik:2014bta,Delduc:2015xdm} mentioned in Footnote \ref{biyb_foot}, and several multi-parameter classes of theories which combine the above with other deformations \cite{Delduc:2017fib,Delduc:2018xug,Seibold:2019dvf}. However, one family of PCM-like models which has not yet been successfully combined with Yang-Baxter deformations is the so-called auxiliary field sigma models (AFSM), which we describe next.

The AFSM is a proposal for generating deformations of the principal chiral model by coupling the PCM to auxiliary fields, i.e. fields with algebraic equations of motion, which introduces interactions in a specific way which preserves integrability. The first iteration of the AFSM appeared in \cite{Ferko:2024ali} and involved a Lagrangian
\begin{align}\label{first_afsm}
    \mathcal{L} = \frac{1}{2} \tr ( j_+ j_- ) + \tr ( v_+ v_- ) + \tr ( j_+ v_- + v_+ j_- ) + E ( \nu_2 ) \, ,
\end{align}
where
\begin{align}
    \nu_2 = \tr ( v_+ v_+ ) \tr ( v_- v_- ) \, ,
\end{align}
and where $E$ is an arbitrary differentiable function of one variable. The family of theories (\ref{first_afsm}) is a $2d$ analogue of the Ivanov-Zupnik formalism for describing theories of duality-invariant electrodynamics in four spacetime dimensions \cite{Ivanov:2002ab,Ivanov:2003uj}. The underlying reason for the similarity with this $4d$ construction is the following. After integrating out the auxiliary fields in equation (\ref{first_afsm}), one arrives at a Lagrangian which depends on $j_{\pm}$ only through the two combinations $\tr ( j_+ j_- )$ and $\tr ( j_+ j_+ ) \tr ( j_- j_- )$. By a change of variables, such a Lagrangian can be written as a function $\mathcal{L} ( S, P )$ of the two variables
\begin{align}
    S = - \frac{1}{2} \tr ( j_+ j_- ) \, , \qquad P^2 = \frac{1}{4} \left( \tr ( j_+ j_+ ) \tr ( j_- j_- ) - \left( \tr ( j_+ j_- ) \right)^2 \right) \, .
\end{align}
The equation of motion associated with any such Lagrangian $\mathcal{L} ( S, P )$ can be written as $\partial_\alpha \CC{J}^\alpha = 0$, where $\CC{J}^\alpha$ is the Noether current associated with the symmetry under right-multiplication of $g$ by a general group element. This current $\CC{J}_\alpha$ has the property
\begin{align}\label{JJ_nice_intro}
    [ \CC{J}_+ , \CC{J}_- ] = [ j_+ , j_- ] \, ,
\end{align}
if and only if the Lagrangian satisfies the differential equation
\begin{align}\label{duality_invariant}
    \mathcal{L}_S^2 - \frac{2 S}{P} \mathcal{L}_S \mathcal{L}_P - \mathcal{L}_P^2 = 1 \, ,
\end{align}
as noticed (in different notation) in \cite{Borsato:2022tmu}. Equation (\ref{duality_invariant}) is identical to the condition for a Lagrangian describing a $4d$ theory of non-linear electrodynamics to enjoy electric-magnetic duality invariance \cite{Gaillard:1981rj,BialynickiBirula:1984tx}, where in that setting the variables $S$ and $P$ are
\begin{align}
    S = - \frac{1}{4} F^{\mu \nu} F_{\mu \nu} \, , \qquad P = - \frac{1}{4} \widetilde{F}^{\mu \nu} F_{\mu \nu} \, ,
\end{align}
where $F = d A$ is the field strength and $\widetilde{F}$ is its Hodge dual.

Returning to $2d$, the condition (\ref{JJ_nice_intro}) on the current $\CC{J}_{\pm}$ is sufficient to establish that the equations of motion for the model are equivalent to the flatness of the Lax connection
\begin{align}\label{lax}
    \mathfrak{L}_{\pm} = \frac{j_{\pm} \pm z \mathfrak{J}_{\pm}}{1 - z^2} \, .
\end{align}
This demonstrates a close connection between the structure of $4d$ theories of duality-invariant electrodynamics and $2d$ deformations of the PCM which preserve integrability while allowing interaction terms that are functions of $\tr ( j_+ j_- )$ and $\tr ( j_+ j_+ ) \tr ( j_- j_- )$.

The goal of the $4d$ Ivanov-Zupnik construction is to ``trivialize'' the task of solving the partial differential equation (\ref{duality_invariant}) by writing a Lagrangian coupled to auxiliary fields which automatically yields a solution to this PDE when the auxiliary fields have been integrated out. Similarly, the AFSM (\ref{first_afsm}) accomplishes the same result in the $2d$ context: for any interaction function $E ( \nu_2 )$, the elimination of the fields $v_{\pm}$ in (\ref{first_afsm}) yields a Lagrangian $\mathcal{L} ( j_{\pm} ) = \mathcal{L} ( S, P )$ which obeys (\ref{duality_invariant}).\footnote{The converse is also true: any Lagrangian which obeys (\ref{duality_invariant}) can be obtained from an auxiliary field presentation for some choice of interaction function $E$, both in the $4d$ and $2d$ settings.} The analysis of integrability in the AFSM crucially involves a field $\CC{J}_{\pm} = - ( j_{\pm} + 2 v_{\pm} )$ which, upon integrating out the auxiliary fields, reduces to the Noether current $\CC{J}_{\pm}$ associated with right-multiplication that we mentioned above.

The family of original auxiliary field sigma models (\ref{first_afsm}) is equivalent to the collection of all deformations of the PCM by functions of the energy-momentum tensor, such as $\TT$ \cite{Zamolodchikov:2004ce,Cavaglia:2016oda,Smirnov:2016lqw} and root-$\TT$ \cite{Ferko:2022cix,Conti:2022egv,Babaei-Aghbolagh:2022uij,Babaei-Aghbolagh:2022leo}.\footnote{Analogous relationships between Ivanov-Zupnik type auxiliary fields and $\TT$-like deformations were also recently discussed in four and six space-time dimensions \cite{Ferko:2023wyi,Ferko:2024zth}.}
A generalization of this construction, which appeared in \cite{Bielli:2024ach}, promotes the interaction function $E ( \nu_2 )$ to a function $E ( \nu_2, \ldots, \nu_N )$ of several variables, and this larger family of integrable models was argued to include deformations of the PCM by both the stress tensor and higher-spin conserved currents, such as the Smirnov-Zamolodchikov higher-spin deformations of  \cite{Smirnov:2016lqw}. The higher-spin auxiliary field sigma model will be reviewed in Section \ref{sec:AFSM_review}. For simplicity, we will use the same term ``auxiliary field sigma model'' and the same acronym AFSM to refer to either the original theory (\ref{first_afsm}) or its higher-spin generalization which will be presented in equation (\ref{AFSM_defn}).

Three other developments concerning the AFSM are worth mentioning. First, in \cite{Fukushima:2024nxm}, the $2d$ auxiliary field sigma model was realized from a construction in four-dimensional Chern-Simons theory \cite{Costello:2019tri,nekrasov_thesis,Costello:2013zra} coupled to auxiliary fields. This is particularly interesting since there has been considerable recent work demonstrating that the $4d$ Chern-Simons theory plays the role of a sort of ``parent theory'' from which many $2d$ integrable field theories can be derived; see \cite{Lacroix:2021iit} for a pedagogical review. The second result is the extension of auxiliary field deformations to symmetric and semi-symmetric space sigma models \cite{Bielli:2024oif} and to more general $\mathbb{Z}_N$ coset models \cite{Cesaro:2024ipq}. The third development, which was worked out in \cite{Bielli:2024khq,Bielli:2024ach}, concerns the interplay between auxiliary field deformations and non-Abelian T-duality. It was shown that such auxiliary field deformations ``commute'' with T-duality in a certain sense, and that the non-Abelian T-dual of the AFSM is related to the AFSM by the same canonical transformation which relates the undeformed PCM to its non-Abelian T-dual \cite{Alvarez:1994wj,Lozano:1995jx}; this also establishes that the T-dual of the AFSM is classically integrable.

Given that the process of activating auxiliary fields deformations of the PCM interacts in a particularly simple way with T-duality -- and since, as we have reviewed above, Yang-Baxter deformations are closely related to T-duality (in the homogeneous case), while a relation between $\TT$ and TsT was already elaborated in \cite{Araujo:2018rho,Sfondrini:2019smd,Apolo:2019zai,Blair:2020ops} -- one might also expect that the AFSM and the YB deformation can be naturally combined. In this work, we will see that this expectation is indeed borne out. We will present a family of doubly-deformed models which incorporates both (bi-)Yang-Baxter and auxiliary field deformations of the principal chiral model, and we will show that this entire multi-parameter family of theories is (weakly) classically integrable, in the sense that its equations of motion admit a Lax representation. We will see that the Lax connection takes the same form for any member of this class of models, and we will check explicitly that these theories have the correct limiting behavior as either of the deformations is turned off -- that is, the theories reduce to the AFSM and the (bi-)YB deformed PCM in appropriate limits. We will also see that this family of Yang-Baxter deformed auxiliary field sigma models (YB-AFSM) admits an equivalent representation via a field redefinition of the auxiliary fields, which has a similar structure to the field redefinition considered in \cite{Bielli:2024khq,Bielli:2024ach} in the context of non-Abelian T-duality. All of these observations support the conclusion that auxiliary field deformations and Yang-Baxter deformations interact in a natural way, and may help point the way towards a physical interpretation for the entire family of combined deformations.

The structure of this paper is as follows. In Section \ref{sec:review}, we review the individual deformations that we have mentioned above, namely auxiliary field sigma models and the Yang-Baxter and bi-Yang-Baxter deformations of the PCM. In Section \ref{sec:yb}, we combine the AFSM and YB deformations to construct the doubly-deformed YB-AFSM which is the main focus of this work, and discuss the properties of this family of models. Section \ref{sec:lax} then establishes that the YB-AFSM is (weakly) classically integrable by giving a zero-curvature representation for its equations of motion, which is the main result of this work. In Section \ref{sec:bi-YB}, we extend these observations to include bi-Yang-Baxter deformations with auxiliary fields (the bi-YB-AFSM or AF-bi-YB model). Finally, Section \ref{sec:conclusion} concludes and outlines directions for future research. We have collected a few ancillary calculations in appendices, namely the computation of the equations of motion for the YB-AFSM in Appendix \ref{app:eom}, a derivation of a useful constraint arising from the flatness of $j_{\pm}$ in Appendix \ref{app:mc_flat}, and some computations concerning the flatness of the bi-YB-AFSM Lax connection in Appendix \ref{app:biyb}.

\section{Review of Undeformed Limits}\label{sec:review}

Beginning from the principal chiral model (\ref{pcm_defn}), whose target space is a Lie group $G$ with Lie algebra $\mathfrak{g}$, the goal of this work is to construct an integrable deformation of the PCM for each pair $( \mathcal{R}, E )$ consisting of a bilinear operator $\mathcal{R}$ on $\mathfrak{g}$ (satisfying certain properties like anti-symmetry and the modified classical Yang-Baxter equation) along with a function $E ( \nu_2 , \ldots , \nu_N ) $ of several variables. To fix notation, as well as to review some known results which make the present work more self-contained, in this section we will discuss the two undeformed limits of this family of deformations. When a parameter called $\eta$ within our family of models is taken to zero, the effect of the bilinear operator $\mathcal{R}$ is turned off, and the theory reduces to the higher-spin auxiliary field sigma model associated with the interaction function $E ( \nu_2 , \ldots , \nu_N )$. On the other hand, if the interaction function $E$ is set to zero and the auxiliary fields are eliminated using their equations of motion, our models reduce to the Yang-Baxter deformed principal chiral model (YB-PCM).

We now recall the essential properties of each of these two limiting cases in turn.

\subsection{Higher-Spin Auxiliary Field Sigma Model}\label{sec:AFSM_review}

In this section, we will review the construction of higher-spin auxiliary field deformations of the principal chiral model which was performed in \cite{Bielli:2024ach}, generalizing the family of spin-$2$ deformations which were constructed in \cite{Ferko:2024ali}. 

Throughout this work, we consider field theories on a flat two-dimensional worldsheet $\Sigma$ with coordinates $(\sigma, \tau)$ that can be packaged into the light-cone combinations
\begin{align}
    \sigma^{\pm} = \frac{1}{2} \left( \tau \pm \sigma \right) \, .
\end{align}
Concretely, one may take $\Sigma$ to be either the plane $\mathbb{R}^{1,1}$ or the cylinder $S^1 \times \mathbb{R}$. We will always use early Greek letters like $\alpha$, $\beta$ for indices on $\Sigma$. In our conventions, the flat worldsheet metric has components $\eta_{+-} = \eta_{-+} = - 2$ in light-cone coordinates.

As we mentioned in the introduction, the physical degree of freedom in all of our models is a group-valued field $g : \Sigma \to G$. The associated left-invariant Maurer-Cartan form,
\begin{align}
    j = g^{-1} d g \, ,
\end{align}
can be pulled back to the worldsheet to define
\begin{align}
    j_\alpha = g^{-1} \partial_\alpha g \, .
\end{align}
By virtue of its definition, this form satisfies the flatness condition
\begin{align}
    \partial_+ j_- - \partial_- j_+ + [ j_+, j_- ] = 0 \, ,
\end{align}
in light-cone coordinates.

In addition to the physical field $g$, the AFSM features a Lie algebra valued auxiliary field $v_{\pm}$. It will sometimes be convenient to expand a generic quantity $X \in \mathfrak{g}$ in a basis of generators $T_A$ for the Lie algebra, writing
\begin{align}
    X = X^A T_A \, ,
\end{align}
so for instance one has $v_{\pm} = v_{\pm}^A T_A$. We will consistently use capital early Latin letters for indices that label the generators of $T_A$ or components in such an expansion.

We will sometimes also write expressions involving the Killing-Cartan form defined by
\begin{align}
    \gamma_{AB} = \tr ( T_A T_B ) \, ,
\end{align}
which we assume to be invertible with inverse $\gamma^{AB}$. This is equivalent to the condition that the Lie algebra $\mathfrak{g}$ is semi-simple, which we mentioned above that we will always assume. In our conventions, the structure constants for the Lie algebra are defined by the relation
\begin{align}
    [ T_A, T_B ] = \tensor{f}{_A_B^C} T_C \, .
\end{align}
With this preamble, we are now prepared to define the auxiliary field sigma model and discuss its properties. This class of theories is defined by a Lagrangian 
\begin{align}\label{AFSM_defn}
    \mathcal{L}_{\text{AFSM}} = \frac{1}{2} \tr ( j_+ j_- ) + \tr ( v_+ v_- ) + \tr ( j_+ v_- + j_- v_+ ) + E ( \nu_2 , \ldots , \nu_N ) \, .
\end{align}
Here $E$ is an arbitrary differentiable function of the $N - 1$ variables $\nu_k$, defined by
\begin{align}\label{nu_defn_afsm}
    \nu_k = \tr ( v_+^k ) \tr ( v_-^k ) \, , \qquad 2 \leq k \leq N \, .
\end{align}
The quantity $N$ is determined by the number of functionally independent traces that can be constructed from the fields $v_{\pm}$. For instance, if the generators $T_A$ of $\mathfrak{g}$ are represented as $M \times M$ matrices, then the objects $v_{\pm} = v_{\pm}^A T_A$ are traceless $M \times M$ matrices (again, tracelessness follows from the assumption that $\mathfrak{g}$ is semi-simple) and hence only the traces $\tr ( v_{\pm}^k )$ for $k = 2 , \ldots , M$ are functionally independent, as higher traces can be expressed in terms of the lower ones. Therefore, in this example one has $N = M$.

The Lagrangian (\ref{AFSM_defn}) gives rise to one equation of motion from varying the auxiliary field $v_{\pm}$ and one Euler-Lagrange equation from varying the group-valued field $g$. The first of these, the auxiliary field equation of motion, can be written as
\begin{align}\label{higher_spin_aux_eom}
    0 \deq j_{\pm}^A + v_{\pm}^A + \sum_{n = 2}^{N} n \frac{\partial E}{\partial \nu_n} \tr ( v_{\pm}^n ) v_{\mp}^{A_1} \ldots v_{\mp}^{A_{n-1}} \gamma^{AB} \tr ( T_{(B} T_{A_1} \ldots T_{A_{n-1} )} ) \, .
\end{align}
Here we have introduced the notation $\deq$ which indicates two quantities which coincide when the auxiliary field equation of motion is satisfied. See Appendix A of \cite{Bielli:2024ach} for a derivation of (\ref{higher_spin_aux_eom}), or for more details on the properties of the AFSM.

Using a standard identity obeyed by the generators $T_A$ of any semi-simple Lie algebra $\mathfrak{g}$, sometimes known as the generalized Jacobi identity, one can show that (\ref{higher_spin_aux_eom}) implies
\begin{align}\label{fundamental_commutator_identity}
    [ v_{\mp} , j_{\pm} ] \deq [ v_{\pm} , v_{\mp} ] \, .
\end{align}
On the other hand, the Euler-Lagrange equation for $g$ is
\begin{align}\label{j_eom}
    \partial_+ ( j_- + 2 v_- ) + \partial_- ( j_+ + 2 v_+ ) \approx 2 \left( [ v_- , j_+ ] + [ v_+, j_- ] \right) \, ,
\end{align}
where we use the notation $\approx$ to denote an equation that holds when the physical field equation of motion is satisfied. Likewise, we will write $\dapprox$ for a relation that is obeyed when both equations of motion hold. For instance, by combining (\ref{j_eom}) and (\ref{fundamental_commutator_identity}), one may write
\begin{align}\label{j_eom_with_aux}
    \partial_+ ( j_- + 2 v_- ) + \partial_- ( j_+ + 2 v_+ ) \dapprox 0 \, ,
\end{align}
since the two commutators on the right side of (\ref{j_eom}) cancel each other when equation (\ref{fundamental_commutator_identity}) is satisfied. This motivates the definition of the objects
\begin{align}
    \CC{J}_{\pm} = - ( j_{\pm} + 2 v_{\pm} ) \, ,
\end{align}
so that the $g$-field equation of motion may be written as the conservation of $\CC{J}_{\pm}$, assuming that the auxiliary field equation of motion is satisfied:
\begin{align}
    \partial_+ \CC{J}_- + \partial_- \CC{J}_+ \dapprox 0 \, .
\end{align}
Finally, we note that the model (\ref{AFSM_defn}) is classically integrable in the following sense. If one defines the Lax connection
\begin{align}\label{lax_connection}
    \CC{L}_{\pm} = \frac{j_{\pm} \pm z \CC{J}_{\pm}}{1 - z^2} \, ,
\end{align}
where $z \in \mathbb{C}$ is the spectral parameter, then the flatness of $\CC{J}_{\pm}$ is equivalent to the $g$-field equation of motion, assuming that the auxiliary field Euler-Lagrange equation is satisfied. That is, if we write the curvature of $\CC{L}_{\pm}$ as
\begin{align}
    d_{\CC{L}} \CC{L} = \partial_+ \CC{L}_- - \partial_- \CC{L}_+ + [ \CC{L}_+ , \CC{L}_- ] \, ,
\end{align}
where $d_{\CC{L}} = d + \CC{L} \, \wedge \, $ is the exterior covariant derivative associated with the connection $\CC{L}$, then $d_{\CC{L}} \CC{L} \deq 0$ if and only if $\partial_\alpha \CC{J}^\alpha \deq 0$. This result was proved in \cite{Bielli:2024ach} and establishes the zero-curvature representation for the AFSM equation of motion. In that work, it was also shown that an infinite set of conserved charges arising from the monodromy matrix of this Lax connection are also in involution, by demonstrating that the Poisson bracket of the spatial component of the Lax connection takes the Maillet $r$/$s$ form \cite{MAILLET198654,MAILLET1986401}. See the lecture notes \cite{Driezen:2021cpd} or the theses \cite{Lacroix:2018njs,Seibold:2020ouf} for introductions to the Maillet construction.

\subsection{Yang-Baxter Deformation of Principal Chiral Model}\label{sec:yb_review}

We now review the other undeformed limit of the class of models to be introduced in Section \ref{sec:yb}, which is the Yang-Baxter deformation of the ordinary principal chiral model. In order to define this deformation, one first chooses a linear operator $\mathcal{R}$ which acts on the Lie algebra $\mathfrak{g}$ associated with the Lie group $G$. This operator is taken to be independent of the physical field $g$ of the model, and is assumed to satisfy three additional properties:
\begin{enumerate}[label = (\roman*)]
    \item The operator $1 - \eta \mathcal{R}$ is invertible on $\mathfrak{g}$, with an inverse that we denote by
    \begin{align}
        \left( 1 - \eta \mathcal{R} \right)^{-1} = \frac{1}{1 - \eta \mathcal{R}} \, ,
    \end{align}
    for all values of $\eta \in \mathbb{R}$.

    \item The operator $\mathcal{R}$ is antisymmetric with respect to the trace on the Lie algebra, which we express by writing $\mathcal{R}^T = - \mathcal{R}$. Explicitly, this means that
    \begin{align}\label{antisymmetry}
        \tr \left( X \mathcal{R} ( Y ) \right) = - \tr \left( \mathcal{R} ( X ) Y \right) \, ,
    \end{align}
    for all $X, Y \in \mathfrak{g}$.

    \item There exists some $c \in \mathbb{C}$ such that $\mathcal{R}$ satisfies the modified classical Yang-Baxter equation (mCYBE),
    \begin{align}\label{mCYBE}
        [ \mathcal{R} X , \mathcal{R} Y ] - \mathcal{R} \left( [ X , \mathcal{R} Y ] \right) - \mathcal{R} \left( [ \mathcal{R} X , Y ] \right) + c^2 [ X, Y ] = 0 \, , 
    \end{align}
    for any $X, Y \in \mathfrak{g}$.
\end{enumerate}
Given such a field-independent operator $\mathcal{R} : \mathfrak{g} \to \mathfrak{g}$, we then define a field-dependent ``dressed'' operator $\mathcal{R}_g$ which acts as
\begin{align}
    \mathcal{R}_g \left( X \right) = g^{-1} \left( \mathcal{R} \left( g X g^{-1} \right) \right) g \, ,
\end{align}
or in terms of the adjoint action $\mathrm{Ad}_g$,
\begin{align}
    \mathcal{R}_g = \mathrm{Ad}_{g^{-1}} \mathcal{R} \mathrm{Ad}_g \, .
\end{align}
The dressed operator $\mathcal{R}_g$ then inherits the same three properties of $\mathcal{R}$ listed above, namely that $1 - \eta \mathcal{R}_g$ is invertible, that $\mathcal{R}_g^T = - \mathcal{R}_g$, and that $\mathcal{R}_g$ satisfies the mCYBE with the same parameter $c \in \mathbb{C}$.

Given such an operator $\mathcal{R}_g$, we define the Lagrangian of the YB-PCM as
\begin{align}\label{yb_defn}
    \mathcal{L}_{\text{YB-PCM}} = - \frac{1}{2} \tr \left( j_+ \frac{1}{1 - \eta \mathcal{R}_g} j_- \right) \, .
\end{align}
When $\eta = 0$, the Lagrangian (\ref{yb_defn}) reduces to the Lagrangian of the PCM given in equation (\ref{pcm_defn}). The equation of motion which arises from varying the fundamental field $g$ is
\begin{align}\label{yb_pcm_eom}
    \partial_+ \left( \frac{1}{1 - \eta \mathcal{R}_g} j_- \right) + \partial_- \left( \frac{1}{1 + \eta \mathcal{R}_g} j_+ \right) \approx 0 \, ,
\end{align}
which takes the form of a conservation equation for a modified current
\begin{align}\label{capital_J_defn}
    J_{\pm} = \frac{1}{1 \pm \eta \mathcal{R}_g} j_{\pm} \, .
\end{align}
We will not review the derivation of the Euler-Lagrange equation (\ref{yb_pcm_eom}) in detail since it is a special case of the equation of motion for the YB-AFSM, when $E = 0$ and the auxiliary fields have been integrated out, and this more general equation of motion is derived in Appendix \ref{app:eom}. As in Section \ref{sec:AFSM_review}, we use the symbol $\approx$ to indicate that two quantities are equal when the field $g$ is on-shell, i.e. when equation (\ref{yb_pcm_eom}) is satisfied. Thus one has
\begin{align}\label{EOM-YB}
    \partial_+ J_- + \partial_- J_+ \approx 0 \, .
\end{align}
Although the current $J_{\pm}$ is conserved, it is not flat, in the sense that it does not satisfy the Maurer-Cartan identity which is obeyed by the field $j_{\pm}$. However, by taking a particular constant multiple of the current $J_{\pm}$, it is possible to obtain a rescaled current which is flat on-shell. More precisely, if we define
\begin{align}
    \widehat{J}_{\pm} = ( 1 - c^2 \eta^2 ) J_{\pm} \, ,
\end{align}
then one finds that
\begin{align}\label{hatted_flatness}
    \partial_+ \widehat{J}_- - \partial_- \widehat{J}_+ + \left[ \widehat{J}_+ , \widehat{J}_- \right] \approx 0 \, ,
\end{align}
which means that the current $\widehat{J}_{\pm}$ is both conserved and flat on-shell.

This observation makes it straightforward to construct a Lax connection for the YB-PCM, which is given by
\begin{align}\label{yb_pcm_lax}
    \CC{L}_{\pm} = \frac{\widehat{J}_{\pm}}{1 \mp z} \, .
\end{align}
The equations of motion for the YB-PCM are equivalent to the flatness of the Lax connection (\ref{yb_pcm_lax}) for any value of the spectral parameter $z \in \mathbb{C}$. That is, the equation
\begin{align}
    0 = d_{\CC{L}} \CC{L} = \partial_+ \CC{L}_- - \partial_- \CC{L}_+ + [ \CC{L}_+ , \CC{L}_- ] \, , 
\end{align}
is satisfied if and only if the equation of motion (\ref{yb_pcm_eom}) holds. More succinctly,
\begin{align}\label{lax_iff}
    d_{\CC{L}} \CC{L} = 0 \; \iff \; \partial_\alpha J^\alpha = 0 \, .
\end{align}
Using the $\approx$ notation introduced above, the $\impliedby$ direction of the statement (\ref{lax_iff}) can also be expressed as the equation $d_{\CC{L}} \CC{L} \approx 0$.

Let us also point out that, having established that $\widehat{J}_{\pm}$ is both flat and conserved on-shell, one can immediately write down an explicit infinite set of conserved higher-spin currents in the YB-PCM. Consider the quantities
\begin{align}\label{cal_J_defn}
    \mathcal{J}_{k \pm} = \tr \left( \widehat{J}_{\pm}^k \right) \, .
\end{align}
Each object $\mathcal{J}_{k \pm}$ carries $k$ copies of a $\pm$ index and thus transforms in a spin-$k$ representation of the Lorentz group. We now show that all of these quantities are conserved. Combining the conservation and flatness equations for $\widehat{J}_{\pm}$ gives
\begin{align}
    \partial_{\pm} \widehat{J}_{\mp} \pm \frac{1}{2} \big[ \widehat{J}_{\pm} , \widehat{J}_{\mp} \big] \approx 0 \, .
\end{align}
Therefore, 
\begin{align}\label{conservation_higher_spin}
    \partial_{\pm} \mathcal{J}_{k \mp} &= k \tr \left( \widehat{J}_{\mp}^{k-1} \partial_{\pm} \widehat{J}_{\mp} \right) \nonumber \\
    &\approx \mp \frac{k}{2} \tr \left( \widehat{J}_{\mp}^{k-1} \big[ \widehat{J}_{\pm} , \widehat{J}_{\mp} \big] \right) \nonumber \\
    &= 0 \, ,
\end{align}
which establishes the existence of a set of higher-spin conserved currents in the YB-PCM.

\subsection{Bi-Yang-Baxter Deformation of Principal Chiral Model}\label{sec:bi-yb_review}

In this section we review another type of deformation of the PCM, known as bi-Yang-Baxter, first introduced in \cite{Klimcik:2014bta}. We will follow the original derivation, introducing some minor modifications in the conventions so as to match the other sections of this work.  The Lagrangian of the bi-YB model reads
\begin{align}\label{bi-yb_defn}
    \mathcal{L}_{\text{bi-YB-PCM}} = - \frac{1}{2} \tr \left( j_+ \frac{1}{1 - \eta \mathcal{R}_g - \zeta \tilde{\mathcal{R}}} j_- \right) \, ,
\end{align}
and the first thing to notice is that the model can be regarded as a deformation of the YB-deformed PCM considered in the previous section, which is indeed recovered by setting the new parameter $\zeta\rightarrow 0$. A second remark is that the deformation introduced by $\zeta$ is associated to an operator $\tilde{\mathcal{R}}$ which can in principle be different from $\mathcal{R}$, but is still required to be antisymmetric \eqref{antisymmetry} and to satisfy the mCYBE \eqref{mCYBE}, with a potentially different constant $\tilde{c}$ appearing in the last term. Finally, the new deformation is completely independent of the physical field $g$ and breaks the right sector of the isometries of the PCM, which was on the other hand preserved by the YB deformation.

We start by computing the equation of motion for the fundamental field $g$, which reads
\begin{equation}\label{EOM-bi-YB}
\partial_{+}J^{\text{K}}_{-}-\partial_{-}J^{\text{K}}_{+} + \zeta[\tilde{\mathcal{R}}(J^{\text{K}}_{-}),J^{\text{K}}_{+}] + \zeta[J^{\text{K}}_{-},\tilde{\mathcal{R}}(J^{\text{K}}_{+})]=0 \,\, ,
\end{equation}
where 
\begin{equation}\label{JK-bi-YB}
J^{\text{K}}_{\pm} = \mp \frac{1}{1 \pm \eta \mathcal{R}_g \pm \zeta \tilde{\mathcal{R}}} j_{\pm} \, ,
\end{equation}
and the superscript K is a reminder that the above current matches the notation of Klim\v{c}\'{i}k \cite{Klimcik:2014bta}. Notice also that, as it should be, in the limit $\zeta \rightarrow 0$ the currents $J^{\text{K}}_{\pm}$ reduce to the YB-currents $J_{\pm}$ defined in \eqref{capital_J_defn}, 
\begin{equation}
J^{\text{K}}_{+} |_{\zeta = 0} = - J_{+} \, , \qquad \qquad  
J^{\text{K}}_{-} |_{\zeta = 0} = + J_{-} \, , 
\end{equation}
and the bi-YB equation of motion \eqref{EOM-bi-YB}, which takes the form of a deformed flatness condition, reduces to the conservation equation \eqref{EOM-YB}. Proceeding in analogy with the YB-deformed model, and following again the original paper \cite{Klimcik:2014bta}, it is then not hard to show that upon inverting the definition of $J^{\text{K}}$ in favor of $j$, one can rewrite the Maurer-Cartan equation $\partial_{+}j_{-}-\partial_{-}j_{+}+[j_{+},j_{-}]=0$ as the following identity:
\begin{equation}\label{by-YB-flatness}
\partial_{+}J^{\text{K}}_{-}+\partial_{-}J^{\text{K}}_{+}+\zeta[J^{\text{K}}_{-},\tilde{\mathcal{R}}(J^{\text{K}}_{+})]+\zeta[J^{\text{K}}_{+},\tilde{\mathcal{R}}(J^{\text{K}}_{-})]+(1-\eta^2 c^2+\zeta^2\tilde{c}^2)[J^{\text{K}}_{-},J^{\text{K}}_{+}]=0 \,\, .
\end{equation}
To do so one has to exploit antisymmetry of $\mathcal{R}$ and $\tilde{\mathcal{R}}$, as well as the mCYBEs satisfied by these operators
\begin{equation}\label{mCYBEs-R-Rtilde}
\begin{aligned}
[ \mathcal{R} X , \mathcal{R} Y ] - \mathcal{R} \left( [ X , \mathcal{R} Y ] \right) - \mathcal{R} \left( [ \mathcal{R} X , Y ] \right) + c^2 [ X, Y ] &= 0 \, ,
\\
 [ \tilde{\mathcal{R}} X , \tilde{\mathcal{R}} Y ] - \tilde{\mathcal{R}} \left( [ X , \tilde{\mathcal{R}} Y ] \right) - \tilde{\mathcal{R}} \left( [ \tilde{\mathcal{R}} X , Y ] \right) + \tilde{c}^2 [ X, Y ] & = 0 \,\, .
\end{aligned}
\end{equation}
Notice that equation \eqref{by-YB-flatness} reduces to (12) in \cite{Klimcik:2014bta} upon choosing $c^2=\tilde{c}^2=-1$ and identifying the two operators as $\tilde{\mathcal{R}}=\mathcal{R}$. We shall now argue that while the restriction $\tilde{c}^2=-1$ is necessary to show weak integrability of the bi-YB model via Klim\v{c}\'{i}k's argument, no restriction on $c$ is needed, nor the identification of the two operators $\mathcal{R}$, $\tilde{\mathcal{R}}$.

To this aim we follow \cite{Klimcik:2014bta} by defining
\begin{equation}\label{Vpm-klimcik}
V_{\pm}^{\text{K}} = \pm \partial_{\pm} J^{\text{K}}_{\mp} \pm \zeta[J^{\text{K}}_{\mp}, \tilde{\mathcal{R}}(J^{\text{K}}_{\pm})] \pm \frac{1}{2}(1-\eta^2 c^2+\zeta^2\tilde{c}^2)[J^{\text{K}}_{-},J^{\text{K}}_{+}] \,\, ,
\end{equation}
in terms of which the EOM \eqref{EOM-bi-YB} and the flatness \eqref{by-YB-flatness} can respectively be rewritten as
\begin{equation}
V_{+}^{\text{K}}+V_{-}^{\text{K}} = 0 \qquad \text{and} \qquad 
V_{+}^{\text{K}}-V_{-}^{\text{K}} = 0 \,\, .
\end{equation}
The proof of weak integrability provided in \cite{Klimcik:2014bta} then relies on a Lax connection whose flatness condition is entirely rewritten in terms of the above sums and differences of $V_{\pm}^{\text{K}}$, in such a way that the EOM \eqref{EOM-bi-YB} and the flatness \eqref{by-YB-flatness} must be separately imposed. To follow \cite{Klimcik:2014bta} while keeping for the moment $c,\tilde{c}$ unspecified and the two operators $\mathcal{R}$ and $\tilde{\mathcal{R}}$ distinct, we consider the Lax connection
\begin{equation}\label{bi-YB-Lax}
\CC{L}_{\pm} = - \left( \zeta \left(\tilde{\mathcal{R}}-i(is\tilde{c}) \right) + (is\tilde{c}) \frac{2i\zeta \pm (1-\eta^2c^2+\zeta^2\tilde{c}^2)}{1\pm z}\right)J^{\text{K}}_{\pm} \,\, ,
\end{equation}
which, up to an overall minus sign, matches the one of Klim\v{c}\'{i}k upon identifying $\mathcal{R} = \tilde{\mathcal{R}}$ and setting $c^2=\tilde{c}^2= -1$, with $\tilde{c}=-is$ and $s=\pm 1$ such that $s^2=1$. Notice that the overall minus sign has the simple purpose of rewriting the flatness condition with a commutator term of the form $[\CC{L}_{+},\CC{L}_{-}]$, rather than $[\CC{L}_{-},\CC{L}_{+}]$ used in \cite{Klimcik:2014bta}, so as to match the notation in the rest of this paper, ensuring the correct reduction of the Lax to the one of the undeformed PCM \cite{Ferko:2024ali}. We can now proceed in writing down the flatness of the Lax
\begin{equation}
\begin{aligned}
& \partial_{+}\CC{L}_{-}-\partial_{+}\CC{L}_{+}+[\CC{L}_{+},\CC{L}_{-}]=
\\
& =-\zeta \left( \tilde{\mathcal{R}}-i(is\tilde{c}) \right) \left( \partial_{+}J^{\text{K}}_{-}-\partial_{-}J^{\text{K}}_{+} \right)
\\
& \quad -(is\tilde{c})\frac{2i\zeta-(1-\eta^2c^2+\zeta^2\tilde{c}^2)}{1-z}\partial_{+}J^{\text{K}}_{-}+(is\tilde{c})\frac{2i\zeta+(1-\eta^2c^2+\zeta^2\tilde{c}^2)}{1+z}\partial_{-}J^{\text{K}}_{+}
\\
& \quad + \zeta(is\tilde{c}) \left( \frac{i\zeta(1+z)-(1-\eta^2c^2+\zeta^2\tilde{c}^2)}{1-z} \right)[\tilde{\mathcal{R}}(J^{\text{K}}_{+}),J^{\text{K}}_{-}]+
\\
& \quad + \zeta(is\tilde{c}) \left( \frac{i\zeta(1-z)+(1-\eta^2c^2+\zeta^2\tilde{c}^2)}{1+z} \right)[J^{\text{K}}_{+},\tilde{\mathcal{R}}(J^{\text{K}}_{-})]+
\\
& \quad + \zeta^2 \left( [\tilde{\mathcal{R}}(J^{\text{K}}_{+}),\tilde{\mathcal{R}}(J^{\text{K}}_{-})] + \tilde{c}^2[J^{\text{K}}_{+},J^{\text{K}}_{-}] \right)
\\
& \quad + (is\tilde{c})^2 \left( \frac{2i\zeta z (1-\eta^2c^2+\zeta^2\tilde{c}^2) - (1-\eta^2c^2+\zeta^2\tilde{c}^2)^2}{1-z^2} \right)[J^{\text{K}}_{+},J^{\text{K}}_{-}] \,\, ,
\end{aligned}
\end{equation}
and, as done in (23) of \cite{Klimcik:2014bta}, comparing to the combination of $V^{\text{K}}_{\pm}$
\begin{equation}
\begin{aligned}
& - \left( \zeta \left(\tilde{\mathcal{R}}-i(is\tilde{c}) \right) + (is\tilde{c}) \frac{2i\zeta + (1-\eta^2c^2+\zeta^2\tilde{c}^2)}{1+ z}\right)V^{\text{K}}_{-} + 
\\
& \qquad  - \left( \zeta \left(\tilde{\mathcal{R}}-i(is\tilde{c}) \right) + (is\tilde{c}) \frac{2i\zeta - (1-\eta^2c^2+\zeta^2\tilde{c}^2)}{1- z}\right)V^{\text{K}}_{+}
\\
& =-\zeta \left( \tilde{\mathcal{R}}-i(is\tilde{c}) \right) \left( \partial_{+}J^{\text{K}}_{-}-\partial_{-}J^{\text{K}}_{+} \right)
\\
& \quad -(is\tilde{c})\frac{2i\zeta-(1-\eta^2c^2+\zeta^2\tilde{c}^2)}{1-z}\partial_{+}J^{\text{K}}_{-}+(is\tilde{c})\frac{2i\zeta+(1-\eta^2c^2+\zeta^2\tilde{c}^2)}{1+z}\partial_{-}J^{\text{K}}_{+}
\\
& \quad + \zeta(is\tilde{c}) \left( \frac{i\zeta(1+z)-(1-\eta^2c^2+\zeta^2\tilde{c}^2)}{1-z} \right)[\tilde{\mathcal{R}}(J^{\text{K}}_{+}),J^{\text{K}}_{-}]+
\\
& \quad + \zeta(is\tilde{c}) \left( \frac{i\zeta(1-z)+(1-\eta^2c^2+\zeta^2\tilde{c}^2)}{1+z} \right)[J^{\text{K}}_{+},\tilde{\mathcal{R}}(J^{\text{K}}_{-})]+
\\
& \quad + \zeta^2 \tilde{\mathcal{R}} \left( [J^{\text{K}}_{+},\tilde{\mathcal{R}}(J^{\text{K}}_{-})] + [\tilde{\mathcal{R}}(J^{\text{K}}_{+}),J^{\text{K}}_{-}] \right)
\\
& \quad + (is\tilde{c}) \left( \frac{2i\zeta z (1-\eta^2c^2+\zeta^2\tilde{c}^2) - (1-\eta^2c^2+\zeta^2\tilde{c}^2)^2}{1-z^2} \right)[J^{\text{K}}_{+},J^{\text{K}}_{-}] \,\, ,
\end{aligned}
\end{equation}
after using the explicit definitions \eqref{Vpm-klimcik}. From the two expressions above, it is clear that the first four lines are the same, while the fifth line coincides upon using the mCYBE satisfied by $\tilde{\mathcal{R}}$, with no constraint on $\tilde{c}$. A problem arises however on the last line, which exhibits different overall factors $(is\tilde{c})^2$ versus $(is\tilde{c})$. This difference can only be removed for
\begin{equation}
(is\tilde{c})^2\equiv (is\tilde{c}) \qquad \Rightarrow \qquad \tilde{c}\equiv -is \,\, ,
\end{equation}
which indeed shows why the use of Klim\v{c}\'{i}k's argument \cite{Klimcik:2014bta} forces the constant $\tilde{c}$ to satisfy $\tilde{c}^2=-1$, while still allowing $c$ to be unconstrained and the operators $\mathcal{R}$, $\tilde{\mathcal{R}}$ to be distinct.
To the best of our knowledge, weak integrability of the bi-YB model has only been shown for the above restriction on $\tilde{c}$, but it cannot be excluded that this requirement might be removed by choosing another, more complicated, form of the Lax connection. This would likely make the proof of weak integrability more involved, as the argument proposed by Klim\v{c}\'{i}k crucially relies on rewriting the flatness of the Lax in terms of combinations of $V_{\pm}^{\text{K}}$.

In light of the above, we now proceed by setting $\tilde{c}=-is$ which leads to the following expression for the curvature of the Lax connection: 
\begin{align}
& \partial_{+}\CC{L}_{-}-\partial_{+}\CC{L}_{+}+[\CC{L}_{+},\CC{L}_{-}]=
\notag \\
& =  \!-\! \left( \zeta \left(\tilde{\mathcal{R}}\!-\!i \right) \!+\! \frac{2i\zeta + (1-\eta^2c^2-\zeta^2)}{1+ z}\right)V^{\text{K}}_{-}  \!-\! \left( \zeta \left(\tilde{\mathcal{R}}\!-\!i \right) \!+\! \frac{2i\zeta - (1-\eta^2c^2-\zeta^2)}{1- z}\right)V^{\text{K}}_{+}
\notag \\
& = -\zeta \left(\tilde{\mathcal{R}}\!-\!i \right) \left( V_{+}^{\text{k}}\!+\!V_{-}^{\text{k}}\right)- \frac{1}{1-z^2}\left( 2i\zeta\left( V_{+}^{\text{k}}\!+\!V_{-}^{\text{k}}\right)-(1-\eta^2c^2-\zeta^2)\left( V_{+}^{\text{k}}\!-\!V_{-}^{\text{k}}\right) \right)
\notag \\
& \qquad \qquad \qquad \qquad \qquad \,\,\,  - \frac{z}{1-z^2}\left( 2i\zeta\left( V_{+}^{\text{k}}\!-\!V_{-}^{\text{k}}\right)-(1-\eta^2c^2-\zeta^2)\left( V_{+}^{\text{k}}\!+\!V_{-}^{\text{k}}\right) \right) \,\, .
\end{align}
This clearly vanishes for any value of the spectral parameter $z$, provided that the sum and difference of $V^{\text{K}}_{\pm}$ are separately set to zero, namely that the EOM \eqref{EOM-bi-YB} and flatness \eqref{by-YB-flatness} are separately imposed. In section \ref{sec:bi-YB} we will construct the auxiliary field deformation of the bi-YB model, taking into account the above restriction on $\tilde{c}$.

\section{Yang-Baxter Deformed AFSM}\label{sec:yb}

We now study the doubly-deformed models which will be the primary focus of this work. This collection of theories generalizes the higher-spin auxiliary field sigma model, and the Yang-Baxter deformed principal chiral model, in the sense that our models reduce to (\ref{AFSM_defn}) when $\eta = 0$ and to (\ref{yb_defn}) after setting $E = 0$ and integrating out the auxiliary fields, as we will check in Section \ref{sec:reduction}. Let us begin by giving the Lagrangian for this family of theories and studying their basic properties, such as implications of the equations of motion.

\subsection{Definition and Properties of Deformed Models}

Consider the family of the Lagrangians
\begin{align}\label{YBAFSM}
    \mathcal{L}_{\text{YB-AFSM}} &= \frac{1}{2} \tr \left( j_- \frac{1}{1 - \eta \mathcal{R}_g} j_+ \right) - \tr ( v_+ v_- ) + 2 \tr \left( v_- \frac{1}{1 - \eta \mathcal{R}_g} v_+ \right) + \tr ( v_- \frac{1}{1 - \eta \mathcal{R}_g} j_+ ) \nonumber \\
    &\qquad + \tr ( j_- \frac{1}{1 - \eta \mathcal{R}_g} v_+ ) + E ( \nu_2 , \ldots , \nu_N ) \, ,
\end{align}
where the arguments of the interaction function $E$ are
\begin{align}
    \nu_k = \tr \left( v_+^k \right) \tr \left( v_-^k \right) \, ,
\end{align}
which is the same as the definition (\ref{nu_defn_afsm}) of these quantities in the AFSM. 

We also note that the Lagrangian (\ref{YBAFSM}) can be written more compactly as
\begin{align}\label{rewritten_YB_AFSM}
    \mathcal{L}_{\text{YB-AFSM}} = \frac{1}{2} \tr \left( ( j_- + 2 v_- ) \frac{1}{1 - \eta \Rg} ( j_+ + 2 v_+ ) \right) - \tr ( v_+ v_- ) + E ( \nu_2 , \ldots , \nu_N ) \, .
\end{align}
To discuss the features of this class of models, it is convenient to introduce the combinations
\begin{align}\label{frakJ_defn}
    \mathfrak{J}_{\pm} = - \frac{1}{1 \mp \eta \mathcal{R}_g} \left( j_\pm + 2 v_\pm \right) \, ,
\end{align}
as well as the related quantities
\begin{align}\label{J_defn}
    J_{\pm} = - \left( \CC{J}_{\pm} + 2 v_{\pm} \right) \, .
\end{align}
To keep track of various relations that hold when the fields of the YB-AFSM satisfy their equations of motion, we will use notation which is similar to that introduced in Section \ref{sec:review}. We write $\deq$ to indicate equality between two expressions that holds when the Euler-Lagrange equation for the auxiliary field $v_\pm$ is satisfied, as we did for the AFSM in Section \ref{sec:AFSM_review}. Likewise, we write $\approx$ to denote equality that holds when the field $g$ obeys its equation of motion, as in Section \ref{sec:yb_review}. Finally, we write $X \dapprox Y$ if the quantities $X$ and $Y$ coincide when \emph{both} the $g$-field equation of motion and the auxiliary field equation of motion hold.

The Euler-Lagrange equations arising from the Lagrangian (\ref{YBAFSM}) are derived in Appendix \ref{app:eom}. When the field $g$ is on-shell, but we do not assume that the auxiliary field's Euler-Lagrange equation holds, one has the relation
\begin{align}
    \partial_- \CC{J}_+ + \partial_+ \CC{J}_- \approx 2 \left( [ v_- , \CC{J}_+ ] + [ v_+ , \CC{J}_- ] \right) \, .
\end{align}
The equation of motion for the auxiliary field can be written as
\begin{align}\label{aux_eom_body}
    \frac{1}{1 \!\mp\! \eta \mathcal{R}_g} \left( j_{\pm} \!+\! 2 v_{\pm} \right) \!-\! v_{\pm} \!+\! \sum_{n=2}^{N} \frac{\partial E}{\partial \nu_n} n \tr ( v_{\pm}^n ) v_{\mp}^{A_1} \ldots v_{\mp}^{A_{n-1}} T^A \tr ( T_{(A} T_{A_1} \ldots T_{A_{n-1} )} ) \deq 0 \, .
\end{align}
We note that equation (\ref{aux_eom_body}) has a similar form to equation (\ref{higher_spin_aux_eom}), and clearly the former reduces to the latter when $\eta = 0$. As in the context of the AFSM, although the structure of the interaction function term in (\ref{aux_eom_body}) is quite complicated, we will not need to use the full equation (\ref{aux_eom_body}) in any of the following arguments. In fact, we will need only one implication of the auxiliary field equation of motion, which can be proven using the generalized Jacobi identity for semi-simple Lie algebras by an argument entirely analogous to the one we have reviewed in Section \ref{sec:AFSM_review}. When (\ref{aux_eom_body}) is satisfied, one finds that
\begin{align}\label{aux_eom_imp_body}
    [ v_+ , \CC{J}_- ] \deq [ v_- , v_+ ] \, .
\end{align}
Using (\ref{aux_eom_imp_body}), we see that when the auxiliary field equation of motion is satisfied, the $g$-field equation of motion may be written more simply as
\begin{align}\label{CCJ_conservation}
    \partial_+ \CC{J}_- + \partial_- \CC{J}_+ \dapprox 0 \, .
\end{align}
That is, when the auxiliary field Euler-Lagrange equation holds, the equation of motion for the physical field $g$ is simply the conservation of the current $\CC{J}_{\pm}$.

\subsection{Reduction in Undeformed Limits}\label{sec:reduction}

As a consistency check on our proposed family of doubly-deformed models (\ref{YBAFSM}), let us now explicitly verify that they reduce to the two singly-deformed models reviewed in Section \ref{sec:review} -- namely the AFSM and YB-PCM -- in the appropriate limits.

The limit $\eta \to 0$ is more straightforward; performing this replacement in (\ref{YBAFSM}) yields
\begin{align}
    \mathcal{L}_{\text{YB-AFSM}} \Big\vert_{\eta = 0} &= \frac{1}{2} \tr \left( j_- j_+ \right) - \tr ( v_+ v_- ) + 2 \tr \left( v_-  v_+ \right) + \tr ( v_-  j_+ ) + \tr ( j_-  v_+ ) + E ( \nu_2 , \ldots , \nu_N ) \nonumber \\
    &= \frac{1}{2} \tr ( j_+ j_- ) + \tr ( v_+ v_- ) + \tr ( j_+ v_- + j_- v_+ ) + E ( \nu_2 , \ldots , \nu_N ) \nonumber \\
    &= \mathcal{L}_{\text{AFSM}} \, ,
\end{align}
which matches (\ref{AFSM_defn}). 

Let us now consider the other limit, in which the interaction function $E$ is set to zero, and the auxiliary fields are integrated out. When $E = 0$, the auxiliary field equation of motion (\ref{aux_eom_body}) becomes
\begin{align}
    j_{\pm} + 2 v_{\pm} - \left( 1 \mp \eta \mathcal{R}_g \right) v_{\pm} \deq 0 \, ,
\end{align}
whose solution is
\begin{align}\label{aux_E_equals_zero}
    v_{\pm} \deq - \frac{1}{1 \pm \eta \mathcal{R}_g} j_{\pm} \, .
\end{align}
In order to streamline the algebra, it is convenient to define the operator 
\begin{align}\label{M_defn}
    M = 1 - \eta \mathcal{R}_g \, ,
\end{align}
so that $M^T = 1 + \eta \mathcal{R}_g$ and
\begin{align}
    M^{-1} = \frac{1}{1 - \eta \mathcal{R}_g} \, , \qquad M^{-T} = \frac{1}{1 + \eta \mathcal{R}_g} \, ,
\end{align}
where we use the notation $M^{-T} = \left( M^{-1} \right)^T = \left( M^T \right)^{-1}$. 

As $\mathcal{R}_g$ is antisymmetric with respect to the trace, the definition (\ref{M_defn}) takes the form $M = 1 + \mathcal{A}$ for an antisymmetric operator $\mathcal{A}$. Any such operator satisfies various identities, which are derived in Appendix D of \cite{Bielli:2024ach}; we will simply quote the results here, and refer the reader to the original work for more details. One has
\begin{align}\label{second_identity_index_free}
    \left( M^{- 1} \right)^T \cdot M^{-1} = \left( M^{-1} \right)^{(S)} = M^{-1} \cdot \left( M^{- 1} \right)^T  \, ,
\end{align}
where we use the superscripts ${}^{(S)}$ and ${}^{(A)}$ for the symmetric and antisymmetric parts, respectively, of any operator. For instance,
\begin{align}\label{minv_idxfree_defns}
    M^{-1} &= \left( M^{-1} \right)^{(S)} + \left( M^{-1} \right)^{(A)} \, , \nonumber \\
    \left( M^{-1} \right)^{(S)} &= \frac{1}{1 - \mathcal{A}^2} \, , \nonumber \\
    \left( M^{-1} \right)^{(A)} &= - \frac{\mathcal{A}}{1 - \mathcal{A}^2} \, .
\end{align}
Likewise, one finds
\begin{align}\label{partial_fractions}
    \left( M^{-1} \right)^T \cdot M^{-1} \cdot \left( M^{-1} \right)^T = \frac{1}{2} \left( M^{-1} \right)^{(S)} + \frac{1}{2} \left( M^T \right)^{-2} \, .
\end{align}
These identities allow us to simplify various terms in the YB-AFSM Lagrangian (\ref{YBAFSM}), when the auxiliary fields are on-shell. For instance, one has
\begin{align}
    \tr \left( v_- \frac{1}{1 - \eta \mathcal{R}_g} v_+ \right) &= \tr \left( v_- M^{-1} v_+ \right) \nonumber \\
    &\deq \tr \left( M^{-1} j_- M^{-1} M^{-T} j_+ \right) \nonumber \\
    &= \tr \left( j_- M^{-T} M^{-1} M^{-T} j_+ \right) \nonumber \\
    &= \frac{1}{2} \tr \left( j_- \left( \left( M^{-1} \right)^{(S)} + \left( M^T \right)^{-2} \right) j_+ \right) \, ,
\end{align}
where in the last step we used (\ref{partial_fractions}), and again the $\deq$ sign indicates equality when the auxiliary field equation of motion (\ref{aux_E_equals_zero}) is satisfied, now specializing to the case $E = 0$. 

Similarly,
\begin{align}
    \tr \left( v_- \frac{1}{1 - \eta \mathcal{R}_g} j_+ \right) &= \tr \left( v_- M^{-1} j_+ \right) \nonumber \\
    &\deq - \tr \left( M^{-1} j_- M^{-1} j_+ \right) \nonumber \\
    &= - \tr \left( j_- M^{-T} M^{-1} j_+ \right) \nonumber \\
    &= - \tr \left( j_- \left( M^{-1} \right)^{(S)} j_+ \right) \, ,
\end{align}
where we have used (\ref{second_identity_index_free}). An identical manipulation gives
\begin{align}
    \tr \left( j_- \frac{1}{1 - \eta \mathcal{R}_g} v_+ \right) &= \tr \left( j_- M^{-1} v_+ \right) \nonumber \\
    &\deq - \tr \left( j_- M^{-1} M^{-T} j_+ \right) \nonumber \\
    &= - \tr \left( j_- \left( M^{-1} \right)^{(S)} j_+ \right) \, .
\end{align}
Furthermore, note that
\begin{align}
    \tr \left( v_+ v_- \right) &= \tr \left( M^{-T} j_+ M^{-1} j_- \right) \nonumber \\
    &= \tr \left( j_+ M^{-2} j_- \right) \, .
\end{align}
Combining the above results, when $E = 0$ and the auxiliary fields are on-shell, we have
\begin{align}
    \mathcal{L}_{\text{YB-AFSM}} &\deq \frac{1}{2} \tr \left( j_- M^{-1} j_+ \right) - \tr \left( j_+ M^{-2} j_- \right) + \tr \left( j_- \left( \left( M^{-1} \right)^{(S)} + \left( M^T \right)^{-2} \right) j_+ \right) \nonumber \\
    &\qquad - 2 \tr \left( j_- \left( M^{-1} \right)^{(S)} j_+ \right) \nonumber \\
    &= \frac{1}{2} \tr \left( j_- M^{-1} j_+ \right) - \tr \left( j_- \left( M^{-1} \right)^{(S)} j_+ \right) \nonumber \\
    &= - \frac{1}{2} \tr \left( j_+ M^{-1} j_- \right) \nonumber \\
    &= \mathcal{L}_{\text{YB-PCM}} \, ,
\end{align}
which establishes that our models correctly reduce to the Yang-Baxter deformed PCM (\ref{yb_defn}) in the limit where $E = 0$ and the auxiliary fields have been eliminated.

\subsection{Equivalent Presentation of YB-AFSM via Field Redefinition}\label{sec:field_redef}

In the formulation (\ref{YBAFSM}) of the YB-AFSM, the role of the auxiliary fields -- or the interpretation of the interaction function $E$ -- is not especially transparent. In this subsection, we will present a physically equivalent formulation of this family of models which will make it easier to see the connection between these deformations and higher-spin conserved currents, at least to leading order around the Yang-Baxter deformed PCM.

Let us consider the redefinition of the auxiliary fields from $v_{\pm}$ to $\tilde{v}_{\pm}$ defined by
\begin{align}\label{aux_field_redef}
    v_+ = \frac{1}{1 + \eta \mathcal{R}_g} \tilde{v}_+ \, , \qquad v_- = \frac{1}{1 - \eta \Rg}\tilde{v}_-  \, .
\end{align}
In terms of these quantities, the scalars $\nu_k$ are given by
%
%
%
\begin{align}\label{nutilde_defn}
    \nu_k = \tr \Big( \underbrace{ \frac{1}{1 + \eta \mathcal{R}_g } \tilde{v}_+ \ldots \frac{1}{1 + \eta \mathcal{R}_g } \tilde{v}_+}_{k \text{ copies } } \Big) \tr \Big( \underbrace{ \frac{1}{1 - \eta \mathcal{R}_g } \tilde{v}_- \ldots \frac{1}{1 - \eta \mathcal{R}_g } \tilde{v}_-}_{k \text{ copies } } \Big) \, .
\end{align}
Performing the field redefinition (\ref{aux_field_redef}) in the Lagrangian (\ref{YBAFSM}) gives the new Lagrangian
\begin{align}\label{YBAFSM_tilde}
    \widetilde{\mathcal{L}}_{\text{YB-AFSM}} &=  \frac{1}{2} \mathrm{tr} \left( j_- \frac{1}{1 - \eta \mathcal{R}_g} j_+ \right) + \mathrm{tr} \left( \tilde{v}_- \frac{1}{1 - \eta^2 \mathcal{R}_g^2} \tilde{v}_+ \right) \nonumber \\
    &\qquad + \mathrm{tr} \left( j_- \frac{1}{1 - \eta^2 \mathcal{R}_g^2} \tilde{v}_+ + \tilde{v}_- \frac{1}{1 - \eta^2 \mathcal{R}_g^2 } j_+ \right) + E \left( \nu_2 , \ldots , \nu_N \right) \, ,
\end{align}
where in deriving (\ref{YBAFSM_tilde}), we have used the identities (\ref{second_identity_index_free}) - (\ref{partial_fractions}) discussed in Section \ref{sec:reduction}.

On the one hand, it is perhaps satisfying that equation (\ref{YBAFSM_tilde}) looks more similar to the Lagrangian (\ref{AFSM_defn}) for the ordinary AFSM. On the other hand, a disadvantage of the presentation (\ref{YBAFSM_tilde}) is that the $g$-field equation of motion will now contain contributions from the interaction function $E$, since the scalars $\nu_k$ defined in equation (\ref{nutilde_defn}) depend on the physical field $g$ through the appearance of $\mathcal{R}_g$ when written in terms of the $\tilde{v}_{\pm}$ variables.

Let us consider the limit of the theories (\ref{YBAFSM_tilde}) when the interaction function $E$ vanishes, as we did for the first formulation of the YB-AFSM in Section \ref{sec:reduction}. In this case, the solution to the auxiliary field equation of motion is simply
\begin{align}\label{tilde_aux_E_equals_zero}
    \tilde{v}_{\pm} \deq - j_{\pm} \, .
\end{align}
This behavior is identical to that of the standard AFSM, where one also has $v_{\pm} \deq - j_{\pm}$ when $E = 0$. Suppose we now consider a small perturbation around this undeformed limit $E = 0$ which involves an interaction function of a single $\nu_k$. That is, we will take
\begin{align}\label{leading_interaction}
    E ( \nu_2 , \ldots , \nu_N ) = \lambda \nu_k
\end{align}
for some fixed $k$, and work to leading order in $\lambda$. The solution (\ref{tilde_aux_E_equals_zero}) for the auxiliary fields is now corrected by terms of order $\lambda$ as
\begin{align}\label{tilde_aux_E_equals_zero_corrected}
    \tilde{v}_{\pm} \deq - j_{\pm} + \mathcal{O} ( \lambda ) \, .
\end{align}
However, since the interaction function $E$ is already of order $\lambda$, when one puts the auxiliary fields on-shell to leading order we simply find
\begin{align}
    E &\deq \lambda \tr \Big( \underbrace{ \frac{1}{1 - \eta \mathcal{R}_g } j_+ \ldots \frac{1}{1 - \eta \mathcal{R}_g } j_+}_{k \text{ copies } } \Big) \tr \Big( \underbrace{ \frac{1}{1 + \eta \mathcal{R}_g } j_- \ldots \frac{1}{1 + \eta \mathcal{R}_g } j_-}_{k \text{ copies } } \Big) + \mathcal{O} ( \lambda^2 ) \nonumber \\
    &= \lambda \tr \left( J_+^k \right) \tr \left( J_-^k \right) + \mathcal{O} ( \lambda^2 ) \nonumber \\
    &= \frac{\lambda}{( 1 - c^2 \eta^2 )^2} \mathcal{J}_{k +} \mathcal{J}_{k -} + \mathcal{O} ( \lambda^2 ) \, ,
\end{align}
where we have recognized the definitions (\ref{capital_J_defn}) and  (\ref{cal_J_defn}) which appeared in the discussion of higher-spin conserved currents in the YB-PCM.

One also finds that, upon substitution of (\ref{tilde_aux_E_equals_zero_corrected}) into (\ref{YBAFSM_tilde}), to first order in $\lambda$ the other terms (besides the interaction function) conspire to reproduce $\mathcal{L}_{\text{YB-PCM}}$, using identities similar to those in Section \ref{sec:reduction}, and one concludes that
\begin{align}\label{first_order_def_currents}
    \widetilde{\mathcal{L}}_{\text{YB-AFSM}} &\deq \mathcal{L}_{\text{YB-PCM}} + \frac{\lambda}{( 1 - c^2 \eta^2 )^2} \mathcal{J}_{k +} \mathcal{J}_{k -} + \mathcal{O} ( \lambda^2 ) \, ,
\end{align}
for the specific choice of interaction function (\ref{leading_interaction}), when the auxiliaries are on-shell and to leading order in $\lambda$.

This makes it clear that, at least to leading order, the effect of introducing our auxiliary field deformations of the Yang-Baxter deformed PCM is to perturb the Lagrangian by a bilinear combination of higher-spin conserved currents.\footnote{Of course, one could have also seen this result using the standard formulation (\ref{YBAFSM}) since they are physically equivalent. The main benefit of the redefined theory $\widetilde{\mathcal{L}}_{\text{YB-AFSM}}$ is that the structure is more similar to that of the standard AFSM, e.g. when $E = 0$ the auxiliary field simply obeys $v_{\pm} \deq - j_{\pm}$.} The combination of currents appearing in the deformation (\ref{first_order_def_currents}) is of particular interest because it takes the Smirnov-Zamolodchikov form, which was shown in \cite{Smirnov:2016lqw} to give rise to a well-defined local operator at the quantum level. More precisely, the coincident point limit
\begin{align}\label{OK_defn}
    \mathcal{O}_k ( x ) = \lim_{y \to x} \left( \mathcal{J}_{k +} ( x ) \mathcal{J}_{k -} ( y ) \right) \, ,
\end{align}
defines a local operator up to total derivative ambiguities in any theory with a pair of higher-spin currents obeying $\partial_{\mp} \mathcal{J}_{k \pm} = 0$. One may therefore deform the quantum theory using this integrated local operator, since on a spacetime without boundary, the spacetime integral eliminates total derivative ambiguities. After performing this deformation, the theory remains integrable and still possesses a set of higher-spin conserved currents, although they will no longer be chirally conserved, instead satisfying an equation of the form $\partial_{\mp} \mathcal{J}_{k \pm} + \partial_{\pm} \mathcal{J}_{ ( k - 2 ) \pm }= 0$. An example of such a deformation is the case $k = 2$, where
\begin{align}
    \mathcal{J}_{\pm \pm} = T_{\pm \pm}
\end{align}
are the components of the energy-momentum tensor in the Yang-Baxter deformed principal chiral model. In this case, the current bilinear deformation corresponds to the $\TT$ deformation \cite{Zamolodchikov:2004ce,Cavaglia:2016oda}; see \cite{Jiang:2019epa} or chapter 2 of \cite{Ferko:2021loo} for reviews.

Although the operator (\ref{OK_defn}) is well-defined quantum-mechanically, the analysis in the present work is entirely classical. As we do not restrict ourselves to deformations that can be defined at the quantum level, we have instead proposed a much larger collection of classical deformed models (\ref{YBAFSM}) which involves \emph{arbitrary} functions of higher-spin combinations of auxiliary fields, rather than only those of the form (\ref{OK_defn}). It is natural to expect that, upon integrating out auxiliary fields, these deformations correspond to deformations of the YB-PCM by arbitrary functions of its higher-spin conserved currents, although we leave an explicit investigation of this conjecture to future work. Instead, we now address the question of whether the deformations (\ref{YBAFSM}) preserve classical integrability.

\section{Integrability of YB-AFSM}\label{sec:lax}

We now show that the Yang-Baxter deformed auxiliary field sigma model is weakly integrable, in the sense that the equation of motion for the physical field $g$ is equivalent to the flatness of a Lax connection, when the auxiliary field equation of motion is satisfied.

In addition to the current $\CC{J}_{\pm}$ discussed above, recall that we have defined the combination $J_{\pm} = - \left( \CC{J}_{\pm} + 2 v_{\pm} \right)$ in equation (\ref{J_defn}). Let us record some identities that $J_{\pm}$ satisfies when the auxiliary field equation of motion is obeyed; the latter, in particular, implies the commutator identity (\ref{aux_eom_imp_body}). From this identity it follows that
\begin{align}\label{JJ_nice}
    [ J_+ , J_- ] &= [ \CC{J}_+ , \CC{J}_- ] + 4 [ v_+ , v_- ] + 2 [ \CC{J}_+ , v_- ] + 2 [ v_+ , \CC{J}_- ] \nonumber \\
    &\deq [ \CC{J}_+ , \CC{J}_- ] + 4 [ v_+ , v_- ] + 2 [ v_- , v_+ ] + 2 [ v_- , v_+ ] \nonumber \\
    &= [ \CC{J}_+ , \CC{J}_- ] \, .
\end{align}
Likewise,
\begin{align}
    [ J_+ , \CC{J}_- ] &= - [ \CC{J}_+ , \CC{J}_- ] - 2 [ v_+, \CC{J}_- ] \nonumber \\
    &\deq - [ \CC{J}_+ , \CC{J}_- ] - 2 [ v_- , v_+ ] \, , 
\end{align}
while on the other hand
\begin{align}
    [ \CC{J}_+ , J_- ] &= - [ \CC{J}_+ , \CC{J}_- ] - 2 [ \CC{J}_+ , v_- ] \nonumber \\
    &\deq - [ \CC{J}_+ , \CC{J}_- ] - 2 [ v_- , v_+ ] \, ,
\end{align}
so we conclude that
\begin{align}\label{JJ_nice_two}
    [ J_+ , \CC{J}_- ] \deq [ \CC{J}_+ , J_- ] \, .
\end{align}
Furthermore, in Appendix \ref{app:mc_flat} we show that -- when both the auxiliary field equation of motion and the $g$-field equation of motion are satisfied, and assuming that $\mathcal{R}_g$ satisfies the modified classical Yang-Baxter equation with parameter $c$ -- one has the relation (\ref{yb_flat}). When written in terms of $J_\pm$, this constraint becomes
\begin{align}\label{J_modified_flatness}
    0 \dapprox \partial_+ J_- - \partial_- J_+ + ( 1 - \eta^2 c^2 ) [J_+ , J_-] \, ,
\end{align}
where we have again used (\ref{JJ_nice}).

The structure of equation (\ref{J_modified_flatness}) is that of a flatness condition for a re-scaled version of the quantity $J_{\pm}$, which is precisely the structure we saw around equation (\ref{hatted_flatness}) in the case of the YB-PCM. Following the notation introduced in that review section, we will introduce ``hatted'' versions of the objects $J_{\pm}$ and $\CC{J}_{\pm}$ as follows:
\begin{align}\label{AFYB-hatted-currents}
    \widehat{J}_{\pm} = ( 1 - c^2 \eta^2 ) J_{\pm} \, , \qquad \widehat{\CC{J}}_{\pm} = ( 1 - c^2 \eta^2 ) \CC{J}_{\pm} \, .
\end{align}
Then equation (\ref{J_modified_flatness}) becomes the flatnes condition
\begin{align}\label{HJ_flatness}
    \partial_+ \widehat{J}_- - \partial_- \widehat{J}_+ + \big[ \widehat{J}_+ , \widehat{J}_- \big] \dapprox 0 \, ,
\end{align}
just as in the case of the Yang-Baxter deformation of the PCM. Furthermore, since the relations (\ref{JJ_nice}) and (\ref{JJ_nice_two}) involving commutators of $J$ and $\CC{J}$ are both quadratic in ``un-hatted'' fields, multiplying each of these equations by the quantity $(1 - c^2 \eta^2)^2$ gives analogous relations for the ``hatted'' fields:
\begin{align}\label{hatted_commutators}
    \big[ \widehat{J}_+ , \widehat{J}_- \big] &\deq \big[ \widehat{\CC{J}}_+ , \widehat{\CC{J}}_- \big] \, , \nonumber \\
    \big[ \widehat{J}_+ , \widehat{\CC{J}}_- \big] &\deq \big[ \widehat{\CC{J}}_+ , \widehat{J}_- \big] \, .
\end{align}
We are now ready to propose the Lax connection for the YB-AFSM, which is
\begin{align}\label{lax_def_body}
    \CC{L}_{\pm} = \frac{ \widehat{J}_{\pm} \pm z \widehat{\CC{J}}_{\pm}}{1 - z^2} \, .
\end{align}
We claim that the flatness of this Lax connection is equivalent to the $g$-field equation of motion, assuming that the auxiliary field Euler-Lagrange equation is satisfied. To see this, let us first compute the commutator
\begin{align}
    [ \CC{L}_+ , \CC{L}_- ] &= \frac{1}{ \left( 1 - z^2 \right)^2 } \Big[ \HJ_+ + z \HCCJ_+ , \HJ_- - z \HCCJ_- \Big] \nonumber \\
    &= \frac{1}{ \left( 1 - z^2 \right)^2 } \left( \big[ \HJ_+ , \HJ_- ] + z \left( \big[ \HCCJ_+ , \HJ_- \big] - \big[ \HJ_+ , \HCCJ_- \big] \right) - z^2 \big[ \HCCJ_+ , \HCCJ_- \big] \right) \nonumber \\
    &\deq \frac{\big[ \HJ_+ , \HJ_- \big]}{1 - z^2} \, ,
\end{align}
where we have used the relations (\ref{hatted_commutators}).

The curvature $d_{\CC{L}} \CC{L}$ of the Lax connection (\ref{lax_def_body}) is then
\begin{align}\label{lax_intermediate}
    d_{\CC{L}} \CC{L} &= \partial_+ \CC{L}_- - \partial_- \CC{L}_+ + [ \CC{L}_+ , \CC{L}_- ] \nonumber \\
    &\deq \frac{1}{1 - z^2} \left( \partial_+ \HJ_- - \partial_- \HJ_+ + \big[ \HJ_+ , \HJ_- \big] - z \left( \partial_+ \HCCJ_- + \partial_- \HCCJ_+  \right) \right) \, .
\end{align}
When the $g$-field equation of motion is satisfied, the first three terms in the parentheses of (\ref{lax_intermediate}) vanish due to the flatness condition (\ref{HJ_flatness}), whereas the terms proportional to $z$ vanish because of the conservation equation (\ref{CCJ_conservation}), which also implies conservation of the rescaled current $\HCCJ_\alpha$. We conclude that
\begin{align}{\label{flat_Lax}}
    d_{\CC{L}} \CC{L} \dapprox 0 \, ,
\end{align}
so the Lax connection is flat when both of the equations of motion are satisfied.

This proves one direction of the ``if and only if'' statement in the definition of weak integrability. To show the other direction, one assumes that $d_{\CC{L}} \CC{L} \deq 0$ for any $z$, which requires that both $\partial_+ \HCCJ_- + \partial_- \HCCJ_+ = 0$ and $\partial_+ \HJ_- - \partial_- \HJ_+ + \big[ \HJ_+ , \HJ_- \big] = 0$ independently. The first condition is the $g$-field equation of motion, and when this equation is satisfied, the second equation automatically holds, establishing the other implication.

\section{Bi-Yang-Baxter Deformed AFSM}\label{sec:bi-YB}

In this section we study the auxiliary field deformation of the bi-Yang-Baxter model introduced in section \ref{sec:bi-yb_review}. The construction will rely on the fact that this new infinite family should reduce to the auxiliary field Yang-Baxter deformation of section \ref{sec:yb} in the limit $\zeta\rightarrow 0$ and to the undeformed bi-YB model of section \ref{sec:bi-yb_review} in the limit where the interaction function vanishes, $E = 0$, and the auxiliary fields are integrated out. These requirements will in turn ensure the correct reduction procedure to the PCM.

\subsection{Definition and Properties}

We begin by presenting the Lagrangian for the bi-YB model, which takes the form
\begin{align}\label{bi-YBAFSM}
    \mathcal{L}_{\text{bi-YB-AFSM}} &= \frac{1}{2} \tr \left( j_- \frac{1}{1 \!-\! \eta \mathcal{R}_g \!-\! \zeta\tilde{\mathcal{R}}} j_+ \right) - \tr ( v_+ v_- ) + 2 \tr \left( v_- \frac{1}{1 \!-\! \eta \mathcal{R}_g \!-\! \zeta\tilde{\mathcal{R}}} v_+ \right) + 
    \notag \\
    &  + \tr ( v_- \frac{1}{1 \!-\! \eta \mathcal{R}_g \!-\! \zeta\tilde{\mathcal{R}}} j_+ ) + \tr ( j_- \frac{1}{1 \!-\! \eta \mathcal{R}_g \!-\! \zeta\tilde{\mathcal{R}}} v_+ ) + E ( \nu_2 , \ldots , \nu_N ) \,\, ,
\end{align}
where the arguments of the interaction function $E$ are once again
\begin{align}
    \nu_k = \tr \left( v_+^k \right) \tr \left( v_-^k \right) \, ,
\end{align}
which is the same as the definition (\ref{nu_defn_afsm}) of these quantities in the AFSM. 

As for the YB deformation, the Lagrangian (\ref{bi-YBAFSM}) can be written more compactly as
\begin{align}\label{rewritten_bi-YB_AFSM}
    \mathcal{L}_{\text{bi-YB-AFSM}} \!=\! \frac{1}{2} \tr \left( ( j_- \!+\! 2 v_- ) \frac{1}{1 \!-\! \eta \Rg \!-\! \zeta\tilde{\mathcal{R}}} ( j_+ \!+\! 2 v_+ ) \right) - \tr ( v_+ v_- ) + E ( \nu_2 , \ldots , \nu_N ) \, ,
\end{align}
and it is convenient to introduce the quantities
\begin{align}\label{bi-YBfrakJ_defn}
    \mathfrak{J}_{\pm}^{\zeta} = - \frac{1}{1 \mp \eta \mathcal{R}_g \mp\zeta \tilde{\mathcal{R}}} \left( j_\pm + 2 v_\pm \right) 
    \qquad \text{and} \qquad
    J_{\pm}^{\zeta} = - \left( \CC{J}_{\pm}^{\zeta} + 2 v_{\pm} \right) \,\, ,
\end{align}
which manifestly reduce to \eqref{frakJ_defn} and \eqref{J_defn} in the limit $\zeta\rightarrow 0$, much as the Lagrangian \eqref{bi-YBAFSM} reduces to \eqref{YBAFSM}. Since the reduction to the AF-YB deformation is clear, and implies the subsequent reduction to the AF-PCM when $\eta \rightarrow 0$, we can directly focus on the reduction to the bi-YB model of section \ref{sec:bi-yb_review} in the limit where $E=0$. To this aim we start by computing the EOM for the auxiliary fields, which take the form
\begin{align}\label{bi-YB-aux_eom_body}
    \frac{1}{1 \!\mp\! \eta \mathcal{R}_g \!\mp\! \zeta\tilde{\mathcal{R}}} \left( j_{\pm} \!+\! 2 v_{\pm} \right) \!-\! v_{\pm} \!+\! \sum_{n=2}^{N} \frac{\partial E}{\partial \nu_n} n \tr ( v_{\pm}^n ) v_{\mp}^{A_1} ... v_{\mp}^{A_{n\!-\!1}} T^A \tr ( T_{\!(A\!} T_{\!A_1\!} ... T_{\!A_{n\!-\!1} )\!} ) \!\deq\! 0 \, .
\end{align}
In the limit $E=0$ it is then clear that the latter equation reduces to
\begin{equation}
\frac{1}{1 \!\mp\! \eta \mathcal{R}_g \!\mp\! \zeta\tilde{\mathcal{R}}} \left( j_{\pm} \!+\! 2 v_{\pm} \right) \!-\! v_{\pm} \deq 0 \qquad \Rightarrow \qquad v_{\pm} \deq -\frac{1}{1 \!\pm\! \eta \mathcal{R}_g \!\pm\! \zeta\tilde{\mathcal{R}}} j_{\pm} = \pm J_{\pm}^{\text{K}} \,\, ,
\end{equation}
which in turn immediately implies that 
\begin{equation}
\mathfrak{J}_{\pm}^{\zeta}\deq -\frac{1}{1 \!\mp\! \eta \mathcal{R}_g \!\mp\! \zeta\tilde{\mathcal{R}}} \left( \frac{1 \!\pm\! \eta \mathcal{R}_g \!\pm\! \zeta\tilde{\mathcal{R}}-2}{1 \!\pm\! \eta \mathcal{R}_g \!\pm\! \zeta\tilde{\mathcal{R}}} \right)j_{\pm} = \frac{1}{1 \!\pm\! \eta \mathcal{R}_g \!\pm\! \zeta\tilde{\mathcal{R}}} j_{\pm} = \mp J_{\pm}^{\text{K}} \,\, ,
\end{equation}
where $J_{\pm}^{\text{K}}$ was defined in \eqref{JK-bi-YB} following the original paper \cite{Klimcik:2014bta} on the undeformed bi-YB model. Using the above equations, the Lagrangian \eqref{rewritten_bi-YB_AFSM} correctly reduces to \eqref{bi-yb_defn},
\begin{equation}
\begin{aligned}
\mathcal{L}_{\text{bi-YB-AFSM}}|_{\text{E}=0} &= \frac{1}{2} \tr \left( ( j_- \!+\! 2 v_- ) \frac{1}{1 \!-\! \eta \Rg \!-\! \zeta\tilde{\mathcal{R}}} ( j_+ \!+\! 2 v_+ ) \right) - \tr ( v_+ v_- ) 
\\
& \deq \frac{1}{2} \tr \left( \left(\frac{1 \!-\! \eta \mathcal{R}_g \!-\! \zeta\tilde{\mathcal{R}}-2}{1 \!-\! \eta \mathcal{R}_g \!-\! \zeta\tilde{\mathcal{R}}}j_{-} \right)\frac{1}{1 \!-\! \eta \mathcal{R}_g \!-\! \zeta\tilde{\mathcal{R}}} \left(\frac{1 \!+\! \eta \mathcal{R}_g \!+\! \zeta\tilde{\mathcal{R}}-2}{1 \!+\! \eta \mathcal{R}_g \!+\! \zeta\tilde{\mathcal{R}}}j_{+} \right)  \right) + 
\\
& \qquad - \tr \left( \left(\frac{1}{1 \!+\! \eta \mathcal{R}_g \!+\! \zeta\tilde{\mathcal{R}}}j_{+} \right) \left(\frac{1}{1 \!-\! \eta \mathcal{R}_g \!-\! \zeta\tilde{\mathcal{R}}}j_{-} \right)  \right) 
\\
& = \frac{1}{2} \tr \left( \left(\frac{1 \!+\! \eta \mathcal{R}_g \!+\! \zeta\tilde{\mathcal{R}}}{1 \!-\! \eta \mathcal{R}_g \!-\! \zeta\tilde{\mathcal{R}}}j_{-} \right) \left(\frac{1 }{1 \!+\! \eta \mathcal{R}_g \!+\! \zeta\tilde{\mathcal{R}}}j_{+} \right)  \right)
\\
& \qquad - \tr \left( \left(\frac{1}{1 \!+\! \eta \mathcal{R}_g \!+\! \zeta\tilde{\mathcal{R}}}j_{+} \right) \left(\frac{1}{1 \!-\! \eta \mathcal{R}_g \!-\! \zeta\tilde{\mathcal{R}}}j_{-} \right)  \right) 
\\
& = \frac{1}{2} \tr \left(  \left(\frac{1 \!+\! \eta \mathcal{R}_g \!+\! \zeta\tilde{\mathcal{R}}-2}{1 \!-\! \eta \mathcal{R}_g \!-\! \zeta\tilde{\mathcal{R}}}j_{-} \right) \left(\frac{1}{1 \!+\! \eta \mathcal{R}_g \!+\! \zeta\tilde{\mathcal{R}}}j_{+} \right)  \right)
\\
& = - \frac{1}{2} \tr \left( j_{+} \frac{1}{1 \!-\! \eta \mathcal{R}_g \!-\! \zeta\tilde{\mathcal{R}}}j_{-}   \right) = \mathcal{L}_{\text{bi-YB-PCM}} \,\, .
\end{aligned}
\end{equation}
We can now proceed in computing the EOM for the physical field $g$, which reads
\begin{equation}\label{AF-bi-YB-EOM}
\partial_{+}\mathfrak{J}^{\zeta}_{-}+\partial_{-}\mathfrak{J}^{\zeta}_{+} + \zeta [\mathfrak{J}^{\zeta}_{+},\tilde{\mathcal{R}}(\mathfrak{J}^{\zeta}_{-})]+\zeta[\tilde{\mathcal{R}}(\mathfrak{J}^{\zeta}_{+}),\mathfrak{J}^{\zeta}_{-}]=0 \,\, .
\end{equation}
Notice that in this form the EOM manifestly reduces to the one obtained for AF-YB, \eqref{CCJ_conservation}, in the limit $\zeta\rightarrow 0$. It exhibits however a structure which is slightly different from the one found by Klim\v{c}\'{i}k, that can be recovered by introducing the new current $L^{\zeta}_{\pm} = \mp \mathfrak{J}^{\zeta}_{\pm}$.
In either case, when $E=0$ and $\mathfrak{J}^{\zeta}_{\pm} \deq \mp J_{\pm}^{\text{K}}$ the undeformed bi-YB EOM \eqref{EOM-bi-YB} is recovered:
\begin{equation}
\partial_{+}J_{-}^{\text{K}}-\partial_{-}J^{\text{K}}_{+}+\zeta[\tilde{\mathcal{R}}(J_{-}^{\text{K}}),J^{\text{K}}_{+}]+\zeta[J_{-}^{\text{K}},\tilde{\mathcal{R}}(J_{+}^{\text{K}})] \deq 0 \,\, .
\end{equation}
Finally, after some manipulation and upon exploiting the YB equations \eqref{mCYBEs-R-Rtilde} satisfied by the operators $\mathcal{R}$ and $\tilde{\mathcal{R}}$, one can rewrite the flatness condition of the current $j$ as
\begin{equation}\label{bi-YB-MC-first-form}
\begin{aligned}
0 \deq &  - \biggl( (\partial_{+}\frak{J}^{\zeta}_{-}-\partial_{-}\frak{J}_{+}^{\zeta})+\zeta[\tilde{\mathcal{R}}(\frak{J}^{\zeta}_{+}),\frak{J}^{\zeta}_{-}]-\zeta[\frak{J}^{\zeta}_{+},\tilde{\mathcal{R}}(\frak{J}^{\zeta}_{-})] -(1-\eta^2c^2+\zeta^2\tilde{c}^2)[\frak{J}^{\zeta}_{+},\frak{J}^{\zeta}_{-}]
\\
& +2(\partial_{+}v_{-}-\partial_{-}v_{+})-2\zeta[v_{+},\tilde{\mathcal{R}}(\frak{J}^{\zeta}_{-})]+2\zeta[\tilde{\mathcal{R}}(\frak{J}^{\zeta}_{+}),v_{-}] \biggr) \, ,
\end{aligned}
\end{equation}
or alternatively, exploiting the second definition in \eqref{bi-YBfrakJ_defn}, as the following mixed condition:
\begin{equation}\label{bi-YB-MC-mixed}
0\deq \partial_{+}J_{-}^{\zeta}-\partial_{-}J_{+}^{\zeta}+\zeta[\tilde{\mathcal{R}}(\mathfrak{J}_{+}^{\zeta}),J_{-}^{\zeta}]-\zeta[J_{+}^{\zeta},\tilde{\mathcal{R}}(\mathfrak{J}_{-}^{\zeta})]+(1-\eta^2c^2+\zeta^2\tilde{c}^2)[J_{+}^{\zeta},J_{-}^{\zeta}] \,\, .
\end{equation}

\subsection{Weak Integrability}
We would now like to show that the EOM of the AF-bi-YB model \eqref{AF-bi-YB-EOM} and the corresponding Maurer-Cartan equation \eqref{bi-YB-MC-mixed} arise from the flatness of a Lax connection, ensuring weak integrability of this infinite family of triply deformed models. As a first attempt, we would like to achieve this exploiting the argument of Klimicik reproduced in section \ref{sec:bi-yb_review}, keeping in mind that the proposed Lax connection should exhibit the correct reduction limits. To this aim, it is useful to recall the form of the Lax connection found for the AFYB models \eqref{lax_def_body}, which relies on the definition of new hatted-currents \eqref{AFYB-hatted-currents}: the appropriate Lax for the AF-bi-YB model should correctly reduce to the latter in the limit $\zeta \rightarrow 0$ and at the same time should reproduce the undeformed bi-YB Lax \eqref{bi-YB-Lax} (with $\tilde{c}=-is$) when $E=0$ and the auxiliary fields are integrated out. We thus begin by defining
\begin{equation}
\widehat{\mathfrak{J}}^{\zeta}_{\pm} = (1-\eta^2c^2-\zeta^2)\frak{J}^{\zeta}_{\pm} 
\qquad \text{and} \qquad
\widehat{J}^{\zeta}_{\pm} = (1-\eta^2c^2-\zeta^2)J_{\pm}^{\zeta} \,\, ,
\end{equation}
which are simple rescaled versions of \eqref{bi-YBfrakJ_defn} and manifestly reduce to \eqref{AFYB-hatted-currents} when $\zeta\rightarrow 0$. In terms of these new quantities, the $g$-field EOM and the MC equation can be written as
\begin{equation}\label{AF-bi-YB-EOM+MC-mixed}
\begin{aligned}
\partial_{+}\widehat{\mathfrak{J}}^{\zeta}_{-}+\partial_{-}\widehat{\mathfrak{J}}^{\zeta}_{+}+\frac{\zeta}{1-\eta^2c^2-\zeta^2} \left( [\widehat{\mathfrak{J}}^{\zeta}_{+},\tilde{\mathcal{R}}(\widehat{\mathfrak{J}}^{\zeta}_{-})]+[\tilde{\mathcal{R}}(\widehat{\mathfrak{J}}^{\zeta}_{+}),\widehat{\mathfrak{J}}^{\zeta}_{-}] 
 \right) &\deq 0 \, ,
\\
\partial_{+}\widehat{J}^{\zeta}_{-}-\partial_{-}\widehat{J}^{\zeta}_{+}+[\widehat{J}^{\zeta}_{+},\widehat{J}^{\zeta}_{-}]+\frac{\zeta}{1-\eta^2c^2-\zeta^2} \left( [\tilde{\mathcal{R}}(\widehat{\frak{J}}^{\zeta}_{+}),\widehat{J}^{\zeta}_{-}]-[\widehat{J}^{\zeta}_{+},\tilde{\mathcal{R}}(\widehat{\frak{J}}^{\zeta}_{-})]
 \right)   & \deq 0 \,\, .
\end{aligned}
\end{equation}
We now propose the following Lax connection:
\begin{equation}\label{AF-bi-YB-proposed-Lax}
\CC{L}_{\pm}=\pm\frac{1}{1-\eta^2c^2-\zeta^2}\left( \zeta(\tilde{\mathcal{R}}-i)+\frac{2i\zeta\pm(1-\eta^2c^2-\zeta^2)}{1\mp z} \right) \left( \frac{\widehat{J}^{\zeta}_{\pm}\pm z \widehat{\frak{J}}^{\zeta}_{\pm}}{1\pm z} \right).
\end{equation}
In the limit $\zeta \rightarrow 0$, this clearly reduces to equation (\ref{lax_def_body}), describing the singly-deformed YB-AFSM. Furthermore, when $E=0$ the auxiliary fields EOM imply that $v_{\pm}\deq \pm J_{\pm}^{\text{K}}$ and $\frak{J}^{\zeta}_{\pm} \deq \mp J^{\text{K}}_{\pm}$, which immediately lead to
$\widehat{J}_{\pm}^{\zeta}\deq \widehat{\frak{J}}^{\zeta}_{\pm}=\mp(1-\eta^2c^2-\zeta^2)J_{\pm}^{\text{K}}$ and in turn to the Lax \eqref{bi-YB-Lax} (with $\tilde{c}=-is$). 

To conclude, by a similar (however more involved) procedure to that of the singly-deformed case, one finds that the flatness condition (\ref{flat_Lax}) remains true for the AF-bi-YB model upon setting $\tilde{c}^2 = -1$. For details, refer to Appendix \ref{app:biyb}.

\section{Conclusion}\label{sec:conclusion}

In this work, we have presented and studied a multi-parameter family of deformations of the principal chiral model. This class of theories combines the (bi-)Yang-Baxter deformation with the recently-proposed higher-spin auxiliary field deformations of the PCM. We have shown that this deformed collection of models admits a zero-curvature representation for its equations of motion, a condition which is sometimes referred to as weak classical integrability. We have also investigated the perturbative interpretation of our auxiliary field deformations to leading order around the Yang-Baxter deformed principal chiral model, using an equivalent presentation of these models involving an auxiliary field redefinition, and thus provided evidence that the role of the auxiliary field couplings is to implement higher-spin deformations of Smirnov-Zamolodchikov type (at least to leading order).

We conclude this work by commenting on a few interesting aspects of our analysis, which may provide directions for further investigation. First, let us point out the structural similarity between the Yang-Baxter deformed AFSM considered here, and the auxiliary field deformations of the non-Abelian T-dual of the PCM considered in \cite{Bielli:2024ach,Bielli:2024khq}. In that context, it was discovered that the process of non-Abelian T-duality ``commutes'' with deformations by auxiliary fields, in a sense which is summarized by the following diagram:
\begin{align}\label{comm_digram}
\begin{tikzcd}[ampersand replacement=\&]
	{\mathcal{L}_{\text{PCM}}} \&\&\& {\mathcal{L}_{\text{TD-PCM}}} \\
	\\
	\&\&\& {\mathcal{L}_{\text{AF-TDSM}}} \\
	{\mathcal{L}_{\text{AFSM}}} \&\& {\mathcal{L}_{\text{TD-AFSM}}}
	\arrow["{\text{T-dualize}}", from=1-1, to=1-4]
	\arrow["{\text{Auxiliaries}}"{description}, from=1-1, to=4-1]
	\arrow["{\text{Auxiliaries}}"{description}, from=1-4, to=3-4]
	\arrow["{\text{T-dualize}}"', from=4-1, to=4-3]
	\arrow["{\substack{\large \text{Field} \\ \normalsize \text{Redefinition}}}"', curve={height=18pt}, tail reversed, from=4-3, to=3-4]
\end{tikzcd} \, .
\end{align}
In the context of that work, the two Lagrangians obtained by traversing the diagram along the two distinct paths bear a close resemblance to the Lagrangians $\mathcal{L}_{\text{YB-AFSM}}$ and $\widetilde{\mathcal{L}}_{\text{YB-AFSM}}$ considered in this work. That is, $\mathcal{L}_{\text{TD-AFSM}}$ takes a form very similar to that of $\mathcal{L}_{\text{YB-AFSM}}$, and $\mathcal{L}_{\text{AF-TDSM}}$ is quite analogous to $\widetilde{\mathcal{L}}_{\text{AFSM}}$; furthermore, the field redefinition indicated in the diagram (\ref{comm_digram}) is essentially identical to the one considered in equation (\ref{aux_field_redef}). The main distinction is that, in this work, the field redefinition involves the operator $M = 1 - \eta \mathcal{R}_g$, while in the context of non-Abelian T-duality, the corresponding operator is $M = 1 + \mathrm{ad}_\Lambda$ where $\Lambda$ is a Lie algebra valued field that is the main degree of freedom in the T-dual model.

In one sense, this result may be expected, since homogeneous Yang-Baxter deformations are closely related to T-duality, so it seems natural that similar structures would emerge in both settings. However, we would like to emphasize that many results in the present article do not sharply distinguish between homogeneous and inhomogeneous Yang-Baxter deformations. All of our analysis of the YB-AFSM -- including the field-redefinition equivalent presentation of Section \ref{sec:field_redef}, the interpretation in terms of higher-spin current bilinear deformations, and the proof of weak classical integrability in Section \ref{sec:lax} -- treat the homogeneous and inhomogeneous cases uniformly, and the arguments are essentially identical in the two contexts, except for occasional appearances of the parameter $c$.\footnote{However, in the case of the bi-YB-AFSM, we saw that a restriction on $\tilde{c}$ was needed for our arguments.} This suggests that the nature of auxiliary field deformations of the Yang-Baxter deformed PCM is essentially blind to the difference between the homogeneous and inhomogeneous settings, and it is intriguing that the similarity to non-Abelian T-duality persists in both cases.

Another interesting direction concerns the physical interpretation of the combined Yang-Baxter and auxiliary field deformations. As we briefly mentioned above, in the pure homogeneous Yang-Baxter case, it has been shown \cite{Borsato:2021fuy} that the deformed model may be re-interpreted as an undeformed model with twisted boundary conditions along the worldsheet cylinder. The argument proceeds by comparing the Lax (\ref{yb_pcm_lax}) for the Yang-Baxter deformed PCM to the Lax $\CC{L}_{\pm}' = \frac{j_{\pm}'}{1 \mp z}$ for the principal chiral model involving a new field $g'$ with Maurer-Cartan form $j'_{\pm}$. Demanding equality of the Lax connections gives 
%
\begin{align}\label{twisted_compare}
    \frac{1}{1 \pm \eta \mathcal{R}_g} j_{\pm} \overset{!}{=} j'_{\pm} \, ,
\end{align}
where $\overset{!}{=}$ indicates that we demand equality, and we have taken $c = 0$ since we are specializing to the homogeneous case. Assuming that the group-valued field $g$ of the Yang-Baxter deformed PCM obeys a standard periodicity condition $g ( \sigma + 2 \pi ) = g ( \sigma )$ along the worldsheet cylinder and working out the implications of the condition (\ref{twisted_compare}) allows one to show
\begin{align}
    g' ( \sigma + 2 \pi ) = W g' ( \sigma ) 
\end{align}
for a twist function $W$ that was identified in \cite{Borsato:2021fuy}.

It would be very interesting to see whether a similar interpretation applies for the doubly-deformed YB-AFSM models considered in this work, at least in the homogeneous case. A subtlety is that, if one na\"ively repeats the analysis reviewed above for the pure hYB model, one would demand equality of the Lax connections
%
%
%
\begin{align}
    \frac{J_{\pm} \pm z \CC{J}_{\pm}}{1 - z^2} \overset{!}{=} \frac{j'_{\pm}}{1 \mp z} \, ,
\end{align}
which is more involved than the comparison (\ref{twisted_compare}) because the left and right sides involve different dependence on the spectral parameter $z$. Nonetheless, if it is possible to overcome this complication, it may be possible to interpret the general class of auxiliary field deformations in terms of twisted boundary conditions. In special cases, such a result would be in accord with the results of \cite{Hernandez-Chifflet:2019sua}, which likewise interpreted deformations by current bilinears of Smirnov-Zamolodchikov type in terms of modified cylinder boundary conditions.

\section*{Acknowledgements}

We thank Sibylle Driezen, Ben Hoare, and Alessandro Sfondrini for helpful discussions. C.\,F. and G.\,T.-M. are grateful to the participants of the meeting ``Integrability in low-supersymmetry theories,'' held in Trani in 2024 
and funded by the COST Action CA22113 by INFN and by Salento University, for stimulating discussions on topics related to the subject of this work.
D.\,B. is supported by Thailand NSRF via PMU-B, grant number B13F670063.
C.\,F. is supported by U.S. Department of Energy grant DE-SC0009999 and funds from the University of California. 
L.\,S. is supported by a postgraduate scholarship at the University of Queensland.
G.\,T.-M. has been supported by the Australian Research Council (ARC) Future Fellowship FT180100353, ARC Discovery
Project DP240101409, and the Capacity Building Package of the University of Queensland.

\appendix

\section{Derivation of Equations of Motion for YB-AFSM}\label{app:eom}

In this appendix, we will obtain the Euler-Lagrange equations for the Yang-Baxter deformed auxiliary field sigma model. We begin from the form of the Lagrangian given in equation (\ref{rewritten_YB_AFSM}), which we also repeat here for convenience:
\begin{align}\label{rewritten_YB_AFSM_appendix}
    \mathcal{L}_{\text{YB-AFSM}} = \frac{1}{2} \tr \left( ( j_- + 2 v_- ) \frac{1}{1 - \eta \Rg} ( j_+ + 2 v_+ ) \right) - \tr ( v_+ v_- ) + E ( \nu_2 , \ldots , \nu_N ) \, .
\end{align}
The algebraic Euler-Lagrange equation associated with the auxiliary field $v_{\pm}$ is
\begin{align}\label{v_eom_app}
    \frac{1}{1 \pm \eta \mathcal{R}_g} \left( j_{\mp} + 2 v_{\mp} \right) - v_{\mp} + \frac{\partial E}{\partial v_{\pm}} \deq 0 \, ,
\end{align}
where we have used that $\mathcal{R}_g$ is antisymmetric so that $\left( \frac{1}{1 - \eta \mathcal{R}_g } \right)^T = \frac{1}{1 + \eta \mathcal{R}_g}$. We also note that the variation of the interaction function can be written explicitly as
\begin{align}\label{interaction_function_variation}
    \frac{\partial E}{\partial v_{\pm}} &= \sum_{n=2}^{N} \frac{\partial E}{\partial \nu_n} n \tr ( v_{\mp}^n ) v_{\pm}^{A_1} \ldots v_{\pm}^{A_{n-1}} T^A \tr ( T_{(A} T_{A_1} \ldots T_{A_{n-1} )} )  \, .
\end{align}
See Appendix A.1 of \cite{Bielli:2024ach} for details on the derivation of equation (\ref{interaction_function_variation}), which is identical to the corresponding variation in the standard (higher-spin) auxiliary field sigma model. Appendix B of the same work also reviews a mathematical fact concerning the generators $T_A$ of a semi-simple Lie algebra $\mathfrak{g}$ with structure constants $\tensor{f}{_A_B^C}$, which we will simply quote here. For any such Lie algebra, one has the ``generalized Jacobi identity''
\begin{align}\label{generalized_jacobi}
    \sum_{i=1}^{n} \tensor{f}{_C_{A_i}^B} \Tr \left( T_{A_1} \ldots T_{A_{i-1}} T_B T_{A_{i+1}} \ldots T_{A_n} \right) = 0 \, .
\end{align}
In particular, if 
\begin{align}
    M_{A_1 \ldots A_n} = \tr \left( T_{(A_1} \ldots T_{A_n)} \right) \, ,
\end{align}
this identity implies 
\begin{align}\label{gen_jac_imp}
    0 = \tensor{f}{_C_{( A_1}^B} M_{A_{2} \ldots A_{n} ) B} \, .
\end{align}
Using the equation of motion (\ref{v_eom_app}), with the explicit formula (\ref{interaction_function_variation}), along with the result (\ref{gen_jac_imp}), one finds that
\begin{align}\label{aux_eom_imp}
    [ v_{\pm} , \CC{J}_{\mp} ] \deq [ v_{\mp} , v_{\pm} ] \, ,
\end{align}
where $\CC{J}_{\pm}$ is defined in (\ref{frakJ_defn}). This implication of the auxiliary field equation of motion will play an important role in the body of this work.

Next we turn to the equation of motion for the physical group-valued field $g$. It is convenient to first obtain some intermediate results concerning the variation of the operator $\mathcal{R}_g$. Under an infinitesimal fluctuation\footnote{In this work, we will always parameterize a variation of the group-valued field as $g \to g e^{\epsilon} = g ( 1 + \epsilon )$ for $\epsilon \in \mathfrak{g}$. By choosing $\epsilon = g^{-1} \delta g$, one can recover a general variation $g \to g + \delta g$ from this parameterization.} $\delta g = g \epsilon$, we note that for any $X$ one has
\begin{align}\label{delta_Rg_X}
    \delta \left( \mathcal{R}_g X \right) &= \delta \left( g^{-1} \left( \mathcal{R} \left( g X g^{-1} \right) g \right) \right) \nonumber \\
    &= \left( - g^{-1} \delta g g^{-1} \right) \left( \mathcal{R} \left( g X g^{-1} \right) \right) g + g^{-1} \left( \mathcal{R} \left( g X g^{-1} \right) \right) \delta g + g^{-1} \left( \mathcal{R} \left( ( \delta g ) X g^{-1} \right) \right) g \nonumber \\
    &\qquad  + g^{-1} \left( \mathcal{R} \left( g X ( - g^{-1} \delta g g^{-1} ) \right) \right) g + g^{-1} \mathcal{R} \left( g \left( \delta X \right) g^{-1} \right) g \nonumber \\
    &= [ \mathcal{R}_g X , \epsilon ] - \mathcal{R}_g \left( [ X, \epsilon ] \right) + \mathcal{R}_g \left( \delta X \right) \, .
\end{align}
To find the variation of the combination $\frac{1}{1 - \eta \mathcal{R}_g} X$, we use the Taylor series expansion
\begin{align}
    \frac{1}{1 - \eta \mathcal{R}_g} X = \sum_{n=0}^{\infty} \eta^n \mathcal{R}_g^n X \, ,
\end{align}
which when combined with the variation (\ref{delta_Rg_X}) implies that
\begin{align}\label{delta_Minv_X}
    \delta \left( \frac{1}{1 - \eta \mathcal{R}_g} X \right) = \frac{1}{1 - \eta \mathcal{R}_g} \left(  \Big[ \frac{\eta \mathcal{R}_g}{1 - \eta \mathcal{R}_g} X , \epsilon \Big] - \eta \mathcal{R}_g \left( \Big[ \frac{1}{1 - \eta \mathcal{R}_g} X , \epsilon \Big] \right)  + \delta X \right) \, .
\end{align}
On the other hand, under $\delta g = g \epsilon$, the Maurer-Cartan form varies as
\begin{align}
    \delta j_{\pm} = \partial_{\pm} \epsilon + [ j_{\pm} , \epsilon ] \, .
\end{align}
Therefore, the variation of the Lagrangian (\ref{rewritten_YB_AFSM_appendix}) under an infinitesimal right-multiplication of the field $g$ is
\begin{align}
    \delta \mathcal{L}_{\text{YB-AFSM}} &= \frac{1}{2} \tr \Bigg( \left( \partial_- \epsilon + [ j_- , \epsilon ] \right) \frac{1}{1 - \eta \mathcal{R}_g} \left( j_+ + 2 v_+ \right) \nonumber \\
    &\qquad \qquad \qquad + \left( j_- + 2 v_- \right)\delta \left( \frac{1}{1 - \eta \mathcal{R}_g} ( j_+ + 2 v_+ ) \right) \Bigg) \nonumber \\
    &= \frac{1}{2} \tr \Bigg( \left( \partial_- \epsilon + [ j_- , \epsilon ] \right) \frac{1}{1 - \eta \mathcal{R}_g} ( j_+ + 2 v_+ ) \nonumber \\
    &\qquad \qquad + \left( j_- + 2 v_- \right) \frac{1}{1 - \eta \mathcal{R}_g} \Bigg(  \Big[ \frac{\eta \mathcal{R}_g}{1 - \eta \mathcal{R}_g} ( j_+ + 2 v_+ ) , \epsilon \Big] + \partial_+ \epsilon + [ j_+ , \epsilon ]   \nonumber \\
    &\qquad \qquad \qquad \qquad \qquad \qquad \qquad \quad - \eta \mathcal{R}_g \left( \Big[ \frac{1}{1 - \eta \mathcal{R}_g} ( j_+ + 2 v_+ ) , \epsilon \Big] \right)  \Bigg) \Bigg) \, .
\end{align}
where we have used (\ref{delta_Minv_X}) with $X = j_+ + 2 v_+$.

It is convenient to express all operators involving $\mathcal{R}_g$, when they appear acting on expressions involving $\epsilon$, in terms of the transpose of the operator acting on combinations which do not involve $\epsilon$. Simplifying these transposes using antisymmetry of $\mathcal{R}_g$ gives
\begin{align}
    \delta \mathcal{L}_{\text{YB-AFSM}} &= \frac{1}{2} \tr \Bigg( \left( \partial_- \epsilon + [ j_- , \epsilon ] \right) \frac{1}{1 - \eta \mathcal{R}_g} ( j_+ + 2 v_+ ) \nonumber \\
    &\qquad \qquad + \left( \frac{1}{1 + \eta \mathcal{R}_g} \left( j_- + 2 v_- \right) \right)  \left(  \Big[ \frac{\eta \mathcal{R}_g}{1 - \eta \mathcal{R}_g} ( j_+ + 2 v_+ ) , \epsilon \Big] + \partial_+ \epsilon + [ j_+ , \epsilon ] \right) \nonumber \\
    &\qquad \qquad + \left( \frac{\eta \mathcal{R}_g}{1 + \eta \mathcal{R}_g} \left( j_- + 2 v_- \right) \right) \Big[ \frac{1}{1 - \eta \mathcal{R}_g} ( j_+ + 2 v_+ ) , \epsilon \Big] \Bigg) \, .
\end{align}
Moving all instances of $\epsilon$ outside of commutators by employing identities like
\begin{align}
    \tr \left( [ X, Y ] Z \right) = \tr \left( [ Z, X ] Y \right) = \tr \left( [ Y, Z ] X \right) \, ,
\end{align}
we find that
\begin{align}
    \delta \mathcal{L}_{\text{YB-AFSM}} &= \frac{1}{2} \tr \Bigg( \left( \frac{1}{1 - \eta \mathcal{R}_g} ( j_+ + 2 v_+ ) \right) \partial_- \epsilon + \left( \frac{1}{1 + \eta \mathcal{R}_g} \left( j_- + 2 v_- \right) \right)  \partial_+ \epsilon \nonumber \\
    &\qquad + \epsilon \Bigg( \Big[ \frac{1}{1 - \eta \mathcal{R}_g} ( j_+ + 2 v_+ ) , j_- \Big] + \Big[ \frac{1}{1 + \eta \mathcal{R}_g} \left( j_- + 2 v_- \right), \frac{\eta \mathcal{R}_g}{1 - \eta \mathcal{R}_g} ( j_+ + 2 v_+ ) \Big]  \nonumber \\
    &\qquad + \Big[ \frac{1}{1 + \eta \mathcal{R}_g} \left( j_- + 2 v_- \right)  , j_+ , \Big] + \Big[ \frac{\eta \mathcal{R}_g}{1 + \eta \mathcal{R}_g} \left( j_- + 2 v_- \right),  \frac{1}{1 - \eta \mathcal{R}_g} ( j_+ + 2 v_+ ) \Big] \Bigg) \Bigg) \, .
\end{align}
When this variation is performed under the integral in the action, so that we may integrate by parts, we therefore obtain
\begin{align}
    \delta S_{\text{YB-AFSM}} &= - \frac{1}{2} \int d^2 \sigma \, \tr \Bigg( \epsilon \Bigg( \partial_- \left( \frac{1}{1 - \eta \mathcal{R}_g} ( j_+ + 2 v_+ ) \right) + \partial_+ \left( \frac{1}{1 + \eta \mathcal{R}_g} \left( j_- + 2 v_- \right) \right) \nonumber \\
    &\qquad - \Big[ \frac{1}{1 - \eta \mathcal{R}_g} ( j_+ + 2 v_+ ) , j_- \Big] - \Big[ \frac{1}{1 + \eta \mathcal{R}_g} \left( j_- + 2 v_- \right), \frac{\eta \mathcal{R}_g}{1 - \eta \mathcal{R}_g} ( j_+ + 2 v_+ ) \Big]  \nonumber \\
    &\qquad - \Big[ \frac{1}{1 + \eta \mathcal{R}_g} \left( j_- + 2 v_- \right)  , j_+ , \Big] - \Big[ \frac{\eta \mathcal{R}_g}{1 + \eta \mathcal{R}_g} \left( j_- + 2 v_- \right),  \frac{1}{1 - \eta \mathcal{R}_g} ( j_+ + 2 v_+ ) \Big] \Bigg) \Bigg) \, .
\end{align}
In terms of the quantities (\ref{frakJ_defn}), we then see that stationarity of the action is equivalent to
\begin{align}
    \partial_- \CC{J}_+ + \partial_+ \CC{J}_- \approx [ \CC{J}_+ , j_- ] + [ \CC{J}_- , \CC{J}_+ - ( j_+ + 2 v_+ )  ] + [ \CC{J}_- , j_+ ] + [ - \CC{J}_- + j_- + 2 v_- , \CC{J}_+ ] \, ,
\end{align}
where we have used the partial fractions decomposition
\begin{align}
    \frac{\eta \mathcal{R}_g}{1 \pm \eta \mathcal{R}_g} X = \mp \frac{1}{1 \pm \eta \mathcal{R}_g} X \pm X \, .
\end{align}
After canceling terms, we arrive at the equation of motion
\begin{align}
    \partial_- \CC{J}_+ + \partial_+ \CC{J}_- \approx 2 \left( [ v_- , \CC{J}_+ ] - [ \CC{J}_-, v_+ ] \right) \, .
\end{align}
As we noted in equation (\ref{aux_eom_imp}), when the auxiliary field equation of motion is satisfied, we have the relations $[ v_{\pm} , \CC{J}_{\mp} ] \deq [ v_{\mp} , v_{\pm} ]$, which allows us to rewrite the $g$-field equation of motion as the conservation equation
\begin{align}\label{g_eom_on_shell}
    \partial_+ \CC{J}_- + \partial_- \CC{J}_+ \dapprox 0 \, .
\end{align}

\section{Derivation of an Implication of the Maurer-Cartan Identity}\label{app:mc_flat}

We have seen that the equations of motion for the YB-AFSM can be recast as the conservation of the modified currents $\mathfrak{J}_{\pm}$, defined in equation (\ref{frakJ_defn}), when the auxiliary field equation of motion is satisfied. These derived quantities are related to the standard Maurer-Cartan form $j_{\pm}$ by the identities
\begin{align}\label{frakJ_to_j}
    j_+ = - \left( 1 - \eta \Rg \right) \CC{J}_+ - 2 v_+ \, , \qquad j_- = - \left( 1 + \eta \Rg \right) \CC{J}_- - 2 v_- \, .
\end{align}
As we have reviewed, by virtue of the definition of $j_{\pm} = g^{-1} \partial_{\pm} g$, this form satisfies the Maurer-Cartan identity
\begin{align}\label{mc_flatness}
    \partial_+ j_- - \partial_- j_+ + [ j_+ , j_- ] = 0 \, .
\end{align}
In this appendix, we will translate the condition (\ref{mc_flatness}) into an equation involving $\mathfrak{J}_\pm$, following the steps that one typically carries out when studying the Yang-Baxter deformation of the ordinary principal chiral model (see, for instance, Section 4 of \cite{Hoare:2021dix}) for a review). Along the way, we will see why the modified classical Yang-Baxter equation (mCYBE),
\begin{align}\label{mcybe}
    [ \Rg X, \Rg Y ] - \Rg \left( [ X, \Rg Y ] \right) - \Rg \left( [ \Rg X, Y ] \right) + c^2 [ X, Y ] = 0 \, ,
\end{align}
naturally emerges from the calculation. We will assume from the beginning of the derivation that $\mathcal{R}$, and hence the dressed operator $\mathcal{R}_g$, is antisymmetric.

We begin by substituting the relations (\ref{frakJ_to_j}) into the flatness condition (\ref{mc_flatness}), finding
\begin{align}
    0 &= - \partial_+ \left(  \left( 1 + \eta \Rg \right) \CC{J}_- + 2 v_- \right) + \partial_- \left( \left( 1 - \eta \Rg \right) \CC{J}_+ + 2 v_+ \right) \nonumber \\
    &\qquad + [ \left( 1 - \eta \Rg \right) \CC{J}_+ + 2 v_+  , \left( 1 + \eta \Rg \right) \CC{J}_- + 2 v_- ] \, .
\end{align}
Note that, by the definition of $\mathcal{R}_g$, for any quantity $X$ we have
\begin{align}\label{Rg_deriv}
    \partial_\pm \left( \mathcal{R}_g X \right) &= \partial_\pm \left( g^{-1} \mathcal{R} \left( g X g^{-1} \right) g \right) \nonumber \\
    &= ( - g^{-1} ( \partial_\pm g ) g^{-1} ) \mathcal{R} \left( g X g^{-1} \right) g + g^{-1} \mathcal{R} ( g X g^{-1} ) \partial_\pm g + g^{-1} \mathcal{R} \left( ( \partial_\pm g ) X g^{-1} \right) g \nonumber \\
    &\qquad + g^{-1} \mathcal{R} \left( g ( \partial_\pm X ) g^{-1} \right) g + g^{-1} \mathcal{R} \left( - g X g^{-1} ( \partial_\pm g ) g^{-1} \right) g \nonumber \\
    &= - [ g^{-1} \partial_\pm g , \mathcal{R}_g X ] + \mathcal{R}_g \left( \partial_\pm X \right) + \mathcal{R}_g \left( [ g^{-1} \partial_\pm g , X ] \right) \nonumber \\
    &= - [ j_\pm , \mathcal{R}_g X ] + \mathcal{R}_g \left( \partial_{\pm} X \right) + \mathcal{R}_g \left( [ j_\pm , X ] \right) \, .
\end{align}
Using (\ref{Rg_deriv}) to evaluate the derivatives of $\mathcal{R}_g \CC{J}_\pm$ gives
\begin{align}
    0 &= - \partial_+ \CC{J}_- + [ j_+ , \eta \Rg \CC{J}_- ] - \eta \Rg \left( \partial_+ \CC{J}_- \right) - \eta \Rg \left( [ j_+ , \CC{J}_- ] \right) - 2 \partial_+ v_- \nonumber \\
    &\qquad + \partial_- \CC{J}_+ + [ j_- , \eta \Rg \CC{J}_+ ] - \eta \Rg \left( \partial_- \CC{J}_+ \right) - \eta \Rg \left( [ j_- , \CC{J}_+ ] \right) + 2 \partial_- v_+ \nonumber \\
    &\qquad + [ \CC{J}_+ , \CC{J}_- ] + 4 [ v_+ , v_- ] + 2 [ \CC{J}_+ , v_- ] + 2 [ v_+ , \CC{J}_- ] - 2 \eta \left( [ \Rg \CC{J}_+ , v_- ] - [ v_+, \Rg \CC{J}_- ] \right) \nonumber \\
    &\qquad - \eta^2 [ \Rg \CC{J}_+ , \Rg \CC{J}_- ] - \eta [ \Rg \CC{J}_+, \CC{J}_- ] + \eta [ \CC{J}_+ , \Rg \CC{J}_- ] \, .
\end{align}
In the remainder of this computation, we will assume that both the auxiliary field equation of motion and the $g$-field equation of motion are satisfied, which allows us to use the relations (\ref{g_eom_on_shell}) and (\ref{aux_eom_imp}). After using these simplifications, we obtain
\begin{align}\label{appB_intermediate_one}
    0 &\dapprox - \left( \partial_+ \CC{J}_- - \partial_- \CC{J}_+ - [ \CC{J}_+ , \CC{J}_- ] \right) + [ j_+ , \eta \Rg \CC{J}_- ] - \eta \Rg \left( [ j_+ , \CC{J}_- ] \right) \nonumber \\
    &\qquad - 2 \left( \partial_+ v_- - \partial_- v_+ \right) + [ j_- , \eta \Rg \CC{J}_+ ]  - \eta \Rg \left( [ j_- , \CC{J}_+ ] \right)  \nonumber \\
    &\qquad - 2 \eta \left( [ \Rg \CC{J}_+ , v_- ] -  [ v_+, \Rg \CC{J}_- ] \right) - \eta^2 [ \Rg \CC{J}_+ , \Rg \CC{J}_- ] - \eta [ \Rg \CC{J}_+, \CC{J}_- ] + \eta [ \CC{J}_+ , \Rg \CC{J}_- ] \, .
\end{align}
We now replace all instances of $j_\pm$ in equation (\ref{appB_intermediate_one}) using equation (\ref{frakJ_to_j}), which gives
\begin{align}
    0 &\dapprox - \left( \partial_+ \CC{J}_- - \partial_- \CC{J}_+ - [ \CC{J}_+ , \CC{J}_- ] \right) - \eta [ \CC{J}_+ , \Rg \CC{J}_- ] + \eta^2 [ \Rg \CC{J}_+ , \Rg \CC{J}_- ] - 2 \eta [ v_+ , \Rg \CC{J}_- ] \nonumber \\
    &\qquad + \eta \Rg \left( [ \CC{J}_+ , \CC{J}_- ] \right) - \eta^2 \Rg \left( [ \Rg \CC{J}_+ , \CC{J}_- ] \right) + 2 \eta \Rg \left( [ v_+ , \CC{J}_- ] \right) \nonumber \\
    &\qquad - 2 \left( \partial_+ v_- - \partial_- v_+ \right)  - \eta [ \CC{J}_- , \Rg \CC{J}_+ ] - \eta^2 [ \Rg \CC{J}_- , \Rg \CC{J}_+ ] - 2 \eta [ v_- , \Rg \CC{J}_+ ] \nonumber \\
    &\qquad + \eta \Rg \left( [ \CC{J}_- , \CC{J}_+ ] \right) + \eta^2 \Rg \left( [ \Rg \CC{J}_- , \CC{J}_+ ] \right) + 2 \eta \Rg \left( [ v_- , \CC{J}_+ ] \right) \nonumber \\
    &\qquad - 2 \eta [ \Rg \CC{J}_+ , v_- ] + 2  \eta [ v_+ , \Rg \CC{J}_- ] - \eta^2 [ \Rg \CC{J}_+ , \Rg \CC{J}_- ] - \eta [ \Rg \CC{J}_+, \CC{J}_- ] + \eta [ \CC{J}_+ , \Rg \CC{J}_- ] \, .
\end{align}
Again we simplify by using the auxiliary field equation of motion and canceling terms: 
\begin{align}\label{appB_intermediate_two}
    0 &\dapprox - \left( \partial_+ \CC{J}_- - \partial_- \CC{J}_+ - [ \CC{J}_+ , \CC{J}_- ] \right) + \eta^2 \left( [ \Rg \CC{J}_+ , \Rg \CC{J}_- ] - \mathcal{R}_g \left( [ \Rg \CC{J}_+ , \CC{J}_- ] \right) - \Rg \left( [ \CC{J}_+ , \Rg  \CC{J}_- ] \right) \right) \nonumber \\
    &\qquad - 2 \left( \partial_+ v_- - \partial_- v_+  \right) \, .
\end{align}
We see that the term proportional to $\eta^2$ on the first line of (\ref{appB_intermediate_two}) matches the first three terms in the modified classical Yang-Baxter equation (\ref{mcybe}). Assuming that the mCYBE is satisfied, we therefore have
\begin{align}\label{yb_flat}
    0 &\dapprox \left( \partial_+ \CC{J}_- - \partial_- \CC{J}_+ - ( 1 - \eta^2 c^2 ) [ \CC{J}_+ , \CC{J}_- ] \right) + 2 \left( \partial_+ v_- - \partial_- v_+ \right) \, .
\end{align}
This is the identity which we wished to derive.

\section{Bi-YB-AFSM Flatness Condition}\label{app:biyb}

In this appendix, we will present some of the more laborious, however important details in proving flatness of the Lax (\ref{AF-bi-YB-proposed-Lax}), which we repeat here for convenience:
\begin{equation}
\CC{L}_{\pm}=\pm\frac{1}{1-\eta^2c^2-\zeta^2}\left( \zeta(\tilde{\mathcal{R}}-i)+\frac{2i\zeta\pm(1-\eta^2c^2-\zeta^2)}{1\mp z} \right) \left( \frac{\widehat{J}^{\zeta}_{\pm}\pm z \widehat{\frak{J}}^{\zeta}_{\pm}}{1\pm z} \right)
\,.
\end{equation}
As seen in the main body, the flatness condition is precisely that the exterior covariant derivative vanishes, or more precisely
\begin{align}
     d_{\CC{L}} \CC{L} &= \partial_+ \CC{L}_- - \partial_- \CC{L}_+ + [ \CC{L}_+ , \CC{L}_- ] = 0 \, .
\end{align}
With some foresight toward computing the curvature, it is useful to write each term with a common denominator of $z$. Doing this, the three terms in the curvature above can be written as
\begin{align}
    \partial_{+}\mathfrak{L}_{-} &\!=\!  -\frac{1}{(1\!-\! z^2)^2}\bigg(\zeta \tilde{\mathcal{R}}(1\!+\!z)(1\!-\!z^2)(\partial_{+}J_{-}\!-\! z\partial_{+}\mathfrak{J}_{-})+(1\!-\!z^2)C_{-}(z)(\partial_{+}J_{-}\!-\! z\partial_{+}\mathfrak{J}_{-})\bigg)\,,\nonumber
    \\
    \partial_{-}\mathfrak{L}_{+} &\!=\! \frac{1}{(1\!-\! z^2)^2}\bigg(\zeta \tilde{\mathcal{R}}(1\!-\!z)(1\!-\!z^2)(\partial_{-}J_{+}\!+\! z\partial_{-}\mathfrak{J}_{+})+(1\!-\!z^2)C_{+}(z)(\partial_{-}J_{+}\!+\! z\partial_{-}\mathfrak{J}_{+})\bigg)\,,\nonumber\\
    [\mathfrak{L}_{+},\mathfrak{L}_{-}] &\!=\! -\frac{1}{(1-z^2)^2}\bigg((1-z^2)\zeta^2 [\tilde{\mathcal{R}}J_{+}, \tilde{\mathcal{R}}J_{-}]+z(1-z^2)\zeta^2 [\tilde{\mathcal{R}}\mathfrak{J}_{+}, \tilde{\mathcal{R}}J_{-}]\nonumber\\
    &\qquad\qquad\quad\:-z(1-z^2)\zeta^2 [\tilde{\mathcal{R}}J_{+}, \tilde{\mathcal{R}}\mathfrak{J}_{-}]-z^2(1-z^2)\zeta^2 [\tilde{\mathcal{R}}\mathfrak{J}_{+}, \tilde{\mathcal{R}}\mathfrak{J}_{-}]\nonumber\\
    &\qquad\qquad\quad\:+\zeta C_{-}(z)(1-z)[\tilde{\mathcal{R}}J_{+},J_{-}]-z(1-z)\zeta C_{-}(z)[\tilde{\mathcal{R}}J_{+},\mathfrak{J}_{-}]\nonumber\\
    &\qquad\qquad\quad\:+z(1-z)\zeta C_{-}(z)[\tilde{\mathcal{R}}\mathfrak{J}_{+},J_{-}]-z^2(1-z)\zeta C_{-}(z)[\tilde{\mathcal{R}}\mathfrak{J}_{+},\mathfrak{J}_{-}]\nonumber\\
    &\qquad\qquad\quad\:+\zeta C_{+}(z)(1+z)[J_{+},\tilde{\mathcal{R}}J_{-}]-z(1+z)\zeta C_{+}(z)[J_{+},\tilde{\mathcal{R}}\mathfrak{J}_{-}]\nonumber\\
    &\qquad\qquad\quad\:+z(1+z)\zeta C_{+}(z)[ \mathfrak{J}_{+},\tilde{\mathcal{R}}J_{-}]-z^2(1+z)\zeta C_{+}(z)[\mathfrak{J}_{+},\tilde{\mathcal{R}}\mathfrak{J}_{-}]\nonumber\\
    &\qquad\qquad\quad\:+C_{+}(z)C_{-}(z)[J_{+}, J_{-}]-zC_{+}(z)C_{-}(z)[J_{+}, \mathfrak{J}_{-}]\nonumber\\
    &\qquad\qquad\quad\:+zC_{+}(z)C_{-}(z)[\mathfrak{J}_{+}, J_{-}]-z^2C_{+}(z)C_{-}(z)[\mathfrak{J}_{+}, \mathfrak{J}_{-}]\bigg)\,.
\end{align}
Here, we introduce the useful functions
\begin{gather}
    C_{+}(z) = 2i\zeta+(1-c^2\eta^2-\zeta^2)-i\zeta(1- z)\,,\nonumber\\
    C_{-}(z) = 2i\zeta-(1-c^2\eta^2-\zeta^2)-i\zeta(1+ z)\,,
\end{gather}
as well as their constant counterparts
\begin{gather}
    C_{+} = C_{+}(0) = i\zeta+(1-c^2\eta^2-\zeta^2)\,,\quad  C_{-} = C_{-}(0) = i\zeta-(1-c^2\eta^2-\zeta^2)\,.
\end{gather}
At this point, it is necessary to group all the terms in the numerator according to powers in $z$. Schematically, all terms in the curvature have the structure 
\begin{gather}
    d_{\CC{L}} \CC{L}(z) = \frac{1}{(1-z^2)^2}P(z)\,.
\end{gather}
In order for the flatness condition to hold, the different powers 
of $z$ in the function $P(z)$ must vanish independently. We begin by listing the terms independent of $z$,
\begin{align}
   P(0) &= C_{-}\partial_{+}J_{-}+C_{+}\partial_{-}J_{+}+C_{-}C_{+}[J_{+},J_{-}]+\zeta C_{-}[\tilde{\mathcal{R}}J_{+},J_{-}]\nonumber\\
   &+\zeta C_{+} [J_{+},\tilde{\mathcal{R}}J_{-}] +\zeta\tilde{\mathcal{R}}\partial_{+}J_{-}+\zeta \tilde{\mathcal{R}}\partial_{-}J_{+}+\zeta^2[\tilde{\mathcal{R}}J_{+},\tilde{\mathcal{R}}J_{-}]\,.
\end{align}
The first-order terms are
\begin{align}
    \frac{d}{dz}\big(P(z)\big)\big\rvert_{z=0} &= \zeta\tilde{\mathcal{R}}\partial_{+}J_{-}  -i \zeta\partial_{+}J_{-} -C_{-}\partial_{+}\mathfrak{J}_{-}-\zeta\tilde{\mathcal{R}}\partial_{+}\mathfrak{J}_{-} - \zeta\tilde{\mathcal{R}} \partial_{-}J_{+}+i\zeta \partial_{-}J_{+} \nonumber\\
     &-C_{-} C_{+}[J_{+},\mathfrak{J}_{-}]+C_{-} C_{+} [\mathfrak{J}_{+},J_{-}]-\zeta C_{-} [\tilde{\mathcal{R}}J_{+},J_{-}] -\zeta C_{-} [\tilde{\mathcal{R}}J_{+},\mathfrak{J}_{-}] \nonumber\\
     &+\zeta C_{-} [\tilde{\mathcal{R}}\mathfrak{J}_{+},J_{-}]+i\zeta C_{-} [J_{+},J_{-}] +C_{+} \partial_{-}\mathfrak{J}_{+}+\zeta C_{+} [J_{+},\tilde{\mathcal{R}}J_{-}] \nonumber\\
     &-\zeta C_{+} [J_{+},\tilde{\mathcal{R}}\mathfrak{J}_{-}] +\zeta C_{+} [\mathfrak{J}_{+},\tilde{\mathcal{R}}J_{-}] -i\zeta C_{+} [J_{+},J_{-}]+\zeta\tilde{\mathcal{R}} \partial_{-}\mathfrak{J}_{+}\nonumber\\
     &+\zeta^2[\tilde{\mathcal{R}}\mathfrak{J}_{+},\tilde{\mathcal{R}}J_{-}] -\zeta^2[\tilde{\mathcal{R}}J_{+},\tilde{\mathcal{R}}\mathfrak{J}_{-}] -i\zeta^2 [\tilde{\mathcal{R}}J_{+},J_{-}] +i\zeta^2 [J_{+},\tilde{\mathcal{R}}J_{-}]\,.
\end{align}
The second order terms are
\begin{align}
    \frac{1}{2!}\frac{d^2}{dz^2}\big(P(z)\big)\big\rvert_{z=0} &= -  C_{-}\partial_{+}J_{-}- \zeta\tilde{\mathcal{R}}\partial_{+}J_{-}  - \zeta\tilde{\mathcal{R}} \partial_{+}\mathfrak{J}_{-}+ i \zeta\partial_{+}\mathfrak{J}_{-}- C_{+}\partial_{-}J_{+} - \zeta\tilde{\mathcal{R}} \partial_{-}J_{+} \nonumber\\
    &- C_{-} C_{+} [\mathfrak{J}_{+},\mathfrak{J}_{-}]+ \zeta C_{-} [\tilde{\mathcal{R}}J_{+},\mathfrak{J}_{-}] - \zeta C_{-} [\tilde{\mathcal{R}}\mathfrak{J}_{+},J_{-}] -  \zeta C_{-} [\tilde{\mathcal{R}}\mathfrak{J}_{+},\mathfrak{J}_{-}]\nonumber\\
    &- i\zeta C_{-} [J_{+},\mathfrak{J}_{-}]+ i\zeta C_{-} [\mathfrak{J}_{+},J_{-}] - \zeta C_{+} [J_{+},\tilde{\mathcal{R}}\mathfrak{J}_{-}] +\zeta  C_{+} [\mathfrak{J}_{+},\tilde{\mathcal{R}}J_{-}] \nonumber\\
    &-\zeta C_{+} [\mathfrak{J}_{+},\tilde{\mathcal{R}}\mathfrak{J}_{-}] + i\zeta C_{+} [J_{+},\mathfrak{J}_{-}]- i\zeta C_{+} [\mathfrak{J}_{+},J_{-}] - \zeta\partial_{-}\mathfrak{J}_{+} \tilde{\mathcal{R}} \nonumber\\
    &+ i\zeta \partial_{-}\mathfrak{J}_{+} - \zeta^2[\tilde{\mathcal{R}}J_{+},\tilde{\mathcal{R}}J_{-}] -\zeta^2 [\tilde{\mathcal{R}}\mathfrak{J}_{+},\tilde{\mathcal{R}}\mathfrak{J}_{-}] + i \zeta^2[\tilde{\mathcal{R}}J_{+},J_{-}] \nonumber\\
    &+ i \zeta^2[\tilde{\mathcal{R}}J_{+},\mathfrak{J}_{-}] - i \zeta^2[\tilde{\mathcal{R}}\mathfrak{J}_{+},J_{-}] + i \zeta^2[J_{+},\tilde{\mathcal{R}}J_{-}] - i \zeta^2[J_{+},\tilde{\mathcal{R}}\mathfrak{J}_{-}] \nonumber\\
    &+ i \zeta^2[\mathfrak{J}_{+},\tilde{\mathcal{R}}J_{-}] + \zeta^2[J_{+},J_{-}]\,.
\end{align}
The third-order terms are
\begin{align}
    \frac{1}{3!}\frac{d^3}{dz^3}\big(P(z)\big)\big\rvert_{z=0} &= -  \zeta\tilde{\mathcal{R}} \partial_{+}J_{-} +  i \zeta\partial_{+}J_{-} + C_{-} \partial_{+}\mathfrak{J}_{-}+  \zeta\tilde{\mathcal{R}} \partial_{+}\mathfrak{J}_{-}+  \zeta\tilde{\mathcal{R}} \partial_{-}J_{+} -  i \zeta\partial_{-}J_{+} \nonumber\\
    &+  \zeta C_{-} [\tilde{\mathcal{R}}\mathfrak{J}_{+},\mathfrak{J}_{-}] -  i \zeta C_{-} [\mathfrak{J}_{+},\mathfrak{J}_{-}] -  C_{+} \partial_{-}\mathfrak{J}_{+}-  \zeta C_{+} [\mathfrak{J}_{+},\tilde{\mathcal{R}}\mathfrak{J}_{-}] \nonumber\\
    &+  i\zeta C_{+} [\mathfrak{J}_{+},\mathfrak{J}_{-}] -\zeta \tilde{\mathcal{R}}  \partial_{-}\mathfrak{J}_{+}-  \zeta^2[\tilde{\mathcal{R}}\mathfrak{J}_{+},\tilde{\mathcal{R}}J_{-}] +  \zeta^2[\tilde{\mathcal{R}}J_{+},\tilde{\mathcal{R}}\mathfrak{J}_{-}] \nonumber\\
    &-  i \zeta^2[\tilde{\mathcal{R}}J_{+},\mathfrak{J}_{-}] +  i\zeta^2 [\tilde{\mathcal{R}}\mathfrak{J}_{+},J_{-}] +  i \zeta^2[\tilde{\mathcal{R}}\mathfrak{J}_{+},\mathfrak{J}_{-}] -  i\zeta^2 [J_{+},\tilde{\mathcal{R}}\mathfrak{J}_{-}] \nonumber\\
    &+  i \zeta^2[\mathfrak{J}_{+},\tilde{\mathcal{R}}J_{-}] -  i \zeta^2[\mathfrak{J}_{+},\tilde{\mathcal{R}}\mathfrak{J}_{-}] - \zeta^2 [J_{+},\mathfrak{J}_{-}]+  \zeta^2[\mathfrak{J}_{+},J_{-}]\,.
\end{align}
Finally, the fourth-order terms are
\begin{align}
    \frac{1}{4!}\frac{d^4}{dz^4}\big(P(z)\big)\big\rvert_{z=0} &=  \zeta\tilde{\mathcal{R}} \partial_{+}\mathfrak{J}_{-}-  i\zeta \partial_{+}\mathfrak{J}_{-}+ \zeta\tilde{\mathcal{R}}  \partial_{-}\mathfrak{J}_{+} -  i\zeta \partial_{-}\mathfrak{J}_{+} +  \zeta^2[\tilde{\mathcal{R}}\mathfrak{J}_{+},\tilde{\mathcal{R}}\mathfrak{J}_{-}] \nonumber\\
    &-  i\zeta^2 [\tilde{\mathcal{R}}\mathfrak{J}_{+},\mathfrak{J}_{-}] -  i \zeta^2[\mathfrak{J}_{+},\tilde{\mathcal{R}}\mathfrak{J}_{-}] -  \zeta^2[\mathfrak{J}_{+},\mathfrak{J}_{-}]\,.
\end{align}
Using various on-shell identities, the Yang-Baxter equations (\ref{mCYBEs-R-Rtilde}) with $\tilde{c}^2=-1$, the equations of motion (\ref{AF-bi-YB-EOM}) and the Bianchi identity (\ref{bi-YB-MC-first-form}), one finds that each order vanishes independently.
\bibliographystyle{utphys}
\bibliography{master}

\end{document}